\documentstyle[12pt,a4wide]{article}
\input epsf
\newcommand{\be}{\begin{equation}}
\newcommand{\ee}{\end{equation}}
\font\mybb=msbm10 at 11pt

\def\bb#1{\hbox{\mybb#1}}

\def\bR {\bb{R}}

\newcommand{\news}{\setcounter{equation}{0}}

\def\ben{\begin{equation}}
\def\een{\end{equation}}
\def\bea{\begin{eqnarray}}
\def\eea{\end{eqnarray}}
\input amssym.def
\input amssym.tex
\begin{document}

\title{\vskip -60pt
\bf \large \bf CENTRAL CONFIGURATIONS IN THREE DIMENSIONS\\[30pt]
\author{Richard A. Battye$^{1}$, 
Gary W.  Gibbons$^{2}$ and Paul M. Sutcliffe$^{3}$\\[10pt]
\\{\normalsize $^{1}$ {\sl Jodrell Bank Observatory, Macclesfield, Cheshire SK11 9DL U.K.}}
\\{\normalsize {\sl $\&$  Department of Physics and Astronomy,
 Schuster Laboratory,}}
\\{\normalsize {\sl University of Manchester, Brunswick St,
 Manchester M13 9PL, U.K.}}
\\{\normalsize {\sl Email : rbattye@jb.man.ac.uk}}\\
\\{\normalsize $^{2}$ {\sl Department of Applied Mathematics and
Theoretical Physics,}}
\\{\normalsize {\sl Centre for Mathematical Sciences, University of Cambridge,}}
\\{\normalsize {\sl Wilberforce Road, Cambridge CB3 0WA, U.K.}}
\\{\normalsize {\sl Email : G.W.Gibbons@damtp.cam.ac.uk}}\\
\\{\normalsize $^{3}$  {\sl Institute of Mathematics,}}
\\{\normalsize {\sl University of Kent at Canterbury,}}\\
{\normalsize {\sl Canterbury, CT2 7NZ, U.K.}}\\
{\normalsize{\sl Email : P.M.Sutcliffe@ukc.ac.uk}}\\}}
\date{October 2002}
\maketitle

\begin{abstract}

We consider the equilibria of point particles under the action of two
body central forces in which there are both repulsive and attractive
interactions, often known as central configurations, with 
diverse applications in physics, in particular as homothetic 
time-dependent 
solutions to Newton's equations of motion and as stationary states in the One
Component Plasma model. Concentrating mainly on the case of an inverse
square law balanced by a linear force, we 
compute numerically equilibria and their statistical  properties. When all the
masses (or charges) of the particles are equal, for small numbers of
points they are regular convex deltahedra, which on
increasing the number of points give way to a multi-shell
structure. In the limit of a large number of points we argue
using an analytic model that they form a  homogeneous spherical
distribution of points, whose spatial distribution appears, from our
preliminary investigation, to be similar to that of a Bernal
hard-sphere liquid. 

\end{abstract}

\vfill \eject

\section{Introduction and summary of results}\news

This is one of a series of papers about central configurations and
related problems involving the equilibria of point particles under the
action of two-body central forces.  The main point of the present work
is to survey what is known mathematically from a wide range of
disciplines and to link this together with some new, mainly numerical,
results of our own, establishing a basis for future work on the
subject.  Our main emphasis here will be on the classical problem of
finding central configurations  of particles associated with an
inverse square  interaction force which are trapped by a linear force,
induced by a harmonic potential.

Such models are very common in a wide variety of physical
applications, but most of our discussion will focus on systems of
gravitating points which in addition to the usual attractive inverse 
square force, experience a repulsive force
proportional to their distance from the origin.
 They arise naturally when seeking
homothetic time-dependent solutions of Newton's equations of motion
for gravitating point particles, which in turn may have some relevance
to Newtonian Cosmology and  models for the large-scale structure of the
universe.

Another physical interpretation arises when the inverse square 
force is thought of as an electrostatic repulsion and the linear force
as an attraction,  due to a uniform background of the opposite
charge. In this guise the problem originally arose in J.J.  Thomson's
static Plum Pudding model of the atom~\cite{JJ1} in which the positive electric
charge is smeared out into a uniform ball (the pudding) while the
negatively charged electrons correspond to the plums. Although
Rutherford's experiments conclusively  demonstrated that this model is
not relevant as a theory of atomic structure, it nevertheless
continues to offer insights into the structure of metals
(with the role of positive and negative charges interchanged)
 and other
condensed matter systems and is often referred to as the One Component
Plasma (OCP) model~\cite{Bau}, or sometimes as classical Jellium.  

Central configurations  are the critical points of a suitable
potential function and those configurations which minimize it are
numerically the easiest to study. In fact almost all of this paper
will be concerned with central configurations which are local minima
that coincide with, or are very close to,  
the absolute minimum of the
potential; only in the case of small numbers of points $(\le 100)$ 
will we claim to
have found the absolute minima. We use two different numerical techniques to
compute these minima.  Firstly, a simple multi-start gradient flow algorithm
which, given a set of random initial conditions, finds the path of
steepest descent toward a local minimum. The other technique is that of
simulated annealing~\cite{sa}, which uses thermal noise to deter the system
from falling into a local minimum which is not the global one.
 We used these two methods in tandem
to increase our confidence in finding the true minimum for small
numbers of points and to find a stationary point close to the true
minimum for larger numbers. By running the codes many
times when the number of points is large,
we were able to deduce that there are very many local minima
with energies close to the absolute minimum. In this
regard it resembles related problems such as that of placing point
charges on a sphere and those of sphere packing.

It turns out that the Plum Pudding interpretation provides the key to
understanding the properties of central configurations for moderate
and large numbers of points and is also quite valuable for understanding
the solutions for small numbers of points. The idea is that for
many purposes one may envisage the equilibria as a packing of
Thomson-type hydrogen atoms, that is, electrically neutral spheres
containing a single negative charge at the centre in a shell of
positive charge.

More quantitatively the spheres correspond to the Thomson atoms
described above. The fact that this correspondence may be elevated to a precise
quantitative tool was apparently first recognized by Leib and
Narnhoffer~\cite{Lieb} who used it to obtain a rigorous lower bound
for the energy of the OCP in terms of a close packing
of Thomson atoms. Our numerical results show that the actual minimum
is incredibly close to the Leib-Narnhoffer bound and leads to a
picture of the equilibria not unlike Bernal's random close packing
model of liquids~\cite{bernal}.
We use the word liquid deliberately because despite
the wide-spread belief that in the limit of infinite numbers of
particles the minimum of the OCP model is given by a
Body Centred Cubic (BCC) 
 crystal, our preliminary results for up to 10,000
particles appear to show no sign of crystallization,
nor long range translational
order. They are, however, crudely consistent with a Bernal liquid. 

A second piece of intuition which appears to be useful  is to consider
points uniformly distributed inside a sphere. Remarkably, by 
using the continuum
limit, an analytic expression can be derived for the probability
distribution for separations in terms of the radius 
of the confining sphere, which is known in
terms of the number of points. This two-point function provides an
analytic test of the homogeneity of the distribution, which is passed
with considerable accuracy. It is
also possible to compute a three-point statistic associated with the distribution of triangles, and we find
agreement there too.

We will present our results for various values of the number of
points, $N$, in three groups designed to exemplify the specific
characteristics of the solutions: 

\begin{itemize}
\item [(I)] Small numbers of points, $N\le 100$ say.

\item [(II)] Moderate numbers of points, say  $100 < N < 1000.$ 

\item [(III)] Large numbers of points.
Here we are able to deal with $1000 \le N \le 10,000$. 
\end{itemize}

\noindent For the most part we will stick to
the case where all the masses (charges) of the particles are equal
($m_1=m_2= .. =m_N=m$).

\medskip
\noindent A summary of the results is as follows :
\medskip 

\noindent In case (I) we claim to have found the absolute minima by
using the two different algorithms with 
a wide range of different initial conditions. For $N\le 12$
the points lie at the vertices of a polyhedron which is a 
deltahedron (one made entirely from triangles) 
except for the antiprism found for $N=8$,
and is regular if $N=4,6$ or $12$. The polyhedron is a tetrahedron
if $N=4$, an octahedron if $N=6$ and an icosohedron if $N=12$. When $N=13$ the
minimum is a single point surrounded by the other twelve in an
icosahedral structure and for $13\le N\le 57$ and $N=60$
there are effectively two shells. There is a link between
$N=13$ being the first value at which a point is found
inside the polyhedron and the fact that at most 12 spheres of equal
radius can touch a given sphere of the same radius. For $58\le N\le 100$
(except $N=60$ which is a particularly symmetric structure) there
are three shells.

\medskip
\noindent In case (II) the configurations found by our algorithms,
which are local minima but may not be the 
absolute minima, look at first glance to be
roughly uniform. However, closer examination of the precise
distribution of points
reveals a clearly defined system of shells. For example, if one plots
the density as a function of radius it oscillates around uniformity
with a regular period. Each of the shells appears to have roughly the
same surface density and the radii of the shells appear to be in
arithmetic progression. This leads to an approximate description of
the number of points in each shell. As the number of points increases
the minimum of the energy comes closer and closer to the lower bound,
suggesting that the assumptions under which it is derived provide a good
picture of the distribution of the particles.

\medskip
\noindent In case (III) we see that a clear spatial uniformity of the
distribution emerges. This is exemplified by computing two-point and
three-point statistics and comparing them to the continuum description
of the problem. With a few minor caveats related to the discreteness
of the distribution, we find remarkable agreement between the analytic
expressions and those found for large $N$; the results for the values
$N=1000$ and $N=10,000$ will be presented. This
uniformity of the density distribution is a  consequence of
Newton's theorem: for an inverse square law, the force due to a
spherically symmetric  distribution of matter is the same as if the
total mass is concentrated at the centre of mass. This is not the
case for any other force law. Of considerable interest is the spatial
distribution of the particles in these uniform distributions. We
computed the distribution of the distance between nearest neighbours
and found it to be sharply peaked, suggesting that each particle can
be thought of as a sphere of fixed radius
and that they may pack as
in the classical sphere packing problem.
 However, a preliminary investigation of
the angular distribution of nearest neighbours reveals no evidence of
long-range orientational order as one might expect, for example, in a
solid. The main caveat to this result is that for large values of $N$
we are unable to have much confidence in having found the global
minimum of the energy. Nonetheless, the asymptotic approach to the lower bound
on the energy suggests that the configurations we have found are very
close to the global minimum.

\section{Central configurations and related problems}

\subsection{Definition of the problem}\news

Classically, central configurations
 are defined as sets of $N$ points ${\bf r}_a \in {\Bbb R}^3$,
$ a=1,\dots,N$ satisfying
\ben
{ \Lambda \over 3} m_a{\bf r}_a + \sum_{b\ne a} {Gm_am_b
({\bf r}_b -{\bf r}_a) \over |{\bf r}  _a -{\bf r}_b |^3 }=0\,, \label{defn1}
\een
where the  constants $G$ and $\Lambda$ are both strictly positive,
for a set of strictly positive masses $(m_1,\dots m_N)$.
The constant $G$ may be thought of as Newton's constant, in which case,
the constant $\Lambda$ has the dimensions of $({\rm time})^{-2}$.
It follows that the centre of mass of the configuration lies at the origin
\ben
\sum_a m_a{\bf r}_a=0\,. \label{com}
\een
It is convenient to divide (\ref{defn1}) by $m_a$ and write it as
\ben
{ \Lambda \over 3} {\bf r}_a + \sum_{b\ne a} {Gm_b
({\bf r}_b -{\bf r}_a) \over |{\bf r}  _a -{\bf r}_b |^3 }=0\,. \label{defn2}
\een

Equation (\ref{defn1}) may be interpreted as stating that each mass
point is in equilibrium under the action of a repulsive radial force
proportional to the mass and the distance from the origin and the
gravitational attraction of the remaining points.  The repulsive force
is such as arises in theories with a cosmological constant
$\Lambda$. It also arises naturally if one makes a time-dependent
homothetic ansatz in Newton's equations of motion. One may instead
think of repulsive Coulomb forces between the particles and an
attraction to the origin. This attraction  can arise from a uniform
density of charge with opposite sign to that of the particles. This
will be discussed in detail later in Section \ref{sec-ocp}.

To begin with we shall show how to eliminate the apparent origin
dependence  and replace the first term by a sum of two-body
repulsions proportional to the separation $r_{ab}= |{\bf r}_a -{\bf r}_b |$.
If we  define the total mass $M$ by
\ben
M=\sum_a m_a\,,
\een
then
\ben
m_a{\bf r}_a=  {1\over M} \sum _{b \ne a} m_a m_b ({\bf r}_a-{\bf r_b})
+ {m_a \over M} \sum_b m_b {\bf r}_b\,.
\label{remove1}
\een

Using (\ref{com}) and (\ref{remove1}) in (\ref{defn1}) we obtain
 \ben \sum _{b \ne a} {\bf F}_{ab}=0\,, \een 
where
 \ben {\bf F}_{ab}= m_a m_b ({\bf r}_a-{\bf r} _b) \left( {\Lambda \over 3M} -{G \over r^3_{ab}  }\right)\,.
\label{twobody} \een 
Note that (\ref{twobody}) is invariant under
translation of the points and, while all solutions of
(\ref{defn1}) are solutions of (\ref{twobody}), these latter solutions can have
any centre of mass. We shall only be interested in solutions
centred on the origin since any solution not centred on the origin
can be obtained from one that is by translation.

Clearly a particular inter-particle distance is picked out, that is,
\ben
r_{ab} = R=\left( {3GM \over \Lambda}\right) ^{ 1\over 3}\,.
\label{distance}
\een
Two particles a distance $R$ apart feel no mutual force.

Note that in units in which $G={\Lambda \over 3}=1$, which we
shall use from now on, and if $m_1=m_2=\dots=m_N=1$, the distance at
which the force vanishes is $R= N^{ 1\over 3}$.
The first few values are
\ben 1.259921\,, \qquad 1.4422496\,, \qquad  1.5874011\,, \qquad 1.7099759\,.
\label{seps}\een

Thus every side of the solutions associated 
with the dipole, triangle and tetrahedron are
given by the first three values respectively. In the case $N=5$ we get
 a triangular bi-pyramid (see Section \ref{sec-N=5}).
 This cannot be regular, but the last value is an estimate for
the average separation. If one believes that the forces essentially
saturate after roughly this distance one gets a close packing model
with diameter roughly $1.7$. In fact, as we shall see later, this
is a slight overestimate and the numerical data suggests the diameter
$d\approx 1.65.$

To gain a further insight into the significance of the radius $R$,
consider  a very large number of points in a roughly spherically
symmetric configuration centred on the origin and in which the
total mass enclosed within a sphere of radius $r$ is $M(r)$. By
Newton's celebrated theorem, the attractive force per unit mass
exerted  on a thin shell of radius $r$ depends only on the masses
enclosed within the shell and is given by 
 \ben {G M(r) \over r^2}\,. \een 
This is an estimate for the second term in (\ref{defn2}).
The cosmic repulsion, i.e. the first term in (\ref{defn2}) is
 \ben {\Lambda \over 3} r\,, \een 
and, therefore, equating these two expressions gives
\ben M(r) = {\Lambda \over 3 G} r^3 \label{density}\,. \een It
follows that any roughly spherically symmetric configuration will
occupy a ball of radius $R$ with roughly uniform density. We shall
see later that for large numbers of points this uniformity holds
with high accuracy.

Note that the argument given above applies only for an inverse
square force law. Thus, we do not expect spatial uniformity for
other force laws and indeed we do not find it to be the case 
(see Section \ref{sec-conc}).

\subsection{Potential functions }

Solutions of (\ref{defn1}) are critical points of
the function
\ben
V=V_{-1} + V_2\,,
\een
where
\ben
V_{-1}= \sum_a \sum_{b <  a} {Gm_am_b  \over |{\bf r}  _a -{\bf r}_b |}\,,
\een
which is homogeneous degree $-1$ and
\ben
V_2=  { \Lambda \over 6} \sum_a m_a {\bf r}^2_a\,,
\een
which is homogeneous degree $2$. Euler's theorem then
gives the virial relation
\ben
V_{-1}=2V_2\,.
\label{virial}
\een
Of course, because the system is rotationally invariant, the critical points
are not isolated, they have 3 rotational zero modes.

One could instead look at critical points of (minus)  the
gravitational potential energy $V_{-1}$ and
 regard $\Lambda \over 3$ as playing the role of a Lagrange multiplier
fixing the value of the function
\ben
I=  { 1\over 2} \sum_a  m_a {\bf r} ^2_a\,.
\een

In what follows we shall refer to solutions as stable if they are
absolute minima of $V$, as metastable if they are local minima and
unstable if the Hessian has some negative eigenvalues. The
terminology is most appropriate for the electrostatic problem since for
the gravitational problem the appropriate potential function is
{\sl minus} $V$. However, the issue of dynamical stability is more
complicated in that case as we shall discuss in detail in our
future paper on the cosmological interpretation of our results.

Finally we remark that, at the expense of introducing three
translational zero modes, one may replace the quadratic
potential $V_2$ by 
\ben {\tilde V}_2= { \Lambda \over 6M}
\sum_{a<b} m_a m_b ({\bf r}_a-{\bf r}_b )^2\,.
\label{manybody1}\een

Thus we need to extremize a sum of two body potentials 
\ben \sum_{a<b} m_a m_b  U(r_{ab})\,,\een 
where 
\ben U(r) = {G \over r} +{\Lambda \over 6M} r^2\,. \label{manybody2}  \een

\subsection{One Component Plasma and Thomson's plum pudding}
\label{sec-ocp}

The One Component Plasma (OCP) \cite{Bau}, sometimes called the
classical Jellium model, is essentially the same problem as
originally studied by Thomson \cite{JJ}  as a model of the atom.
Nowadays it is often used as a model for metals at high density in
which one assumes that quantum mechanically degenerate electrons
provide a uniform background of {\sl negative} charge in which
there are immersed positively charged nuclei. Of course in
Thomson's original model the roles of positive and negative charges
are reversed.

Note that the problem of placing point charges on a sphere (see Section 2.8) is often, but mistakenly, referred to
as the Thomson problem. 
For the Thomson problem (or equivalently the OCP, since the
sign of the charges is irrelevant here)
one considers a uniformly positively
charged domain $\Omega \subset {\Bbb R}^3$ with volume $A$
containing $N$ negatively charged corpuscles. The sum of the
negative charges is taken to be  equal to the total positive charge. The
potential energy of the system is taken to consist of three parts
\ben V_{OCP}= V_{--}+ V_{+-} + V_{++}\,. \een $V_{--}$ is the
positive mutual electrostatic energy of the negatively charged
particles.
 $V_{+-}$ is the
electrostatic potential energy of the negative charges in the
potential generated by the uniformly distributed positive
background.  Finally, one includes the potential energy, $V_{++}$, of the
uniformly charged  positive background. 

Usually one takes all of the charges to have the same value, but
one may consider the case when they differ. If one does so one
obtains a system identical to the one discussed in Sections 2.1 and 2.2.
Rather than introducing further unnecessary notation we
shall continue with our present conventions leaving to the reader
the trivial task of transcription to the  electrostatic units of
his or her choice (ref.~\cite{joke} may prove useful in this respect).

With the proviso that all particles must lie inside $\Omega,$
we have that
 \ben V_{--}= V_{-1}\,, \een
\ben V_{+-}= -G M \sum_a \left( {m_a \over A} \int {d^3 {\bf r}
\over |{\bf r} -{\bf r_a }| } \right)\,,\een 
and 
\ben V_{++} = {G
M^2 \over 2} \left( { 1\over A^2 } \int \int { d^3{\bf r} d^3
 {{\bf r} ^\prime } \over |{\bf r} - { {\bf r} ^\prime} | } \right)\,.
\een

In the case when $\Omega$ is taken to be a ball of radius $R$ we
can evaluate the integrals. 
\ben V_{+-} = -{ 3GM^2 \over 2R} +{ GM
\over 2R^3 } \sum m_a {\bf r}^2 _a\,,
\een
\ben V_{++} =  { 3 \over
5} {G M^2 \over R}\,. \een
Thus \ben V_{OCP} = V_{-1} + { 3GM \over
\Lambda R^3} V_2 - {9 GM^2 \over 10 R}\,. \een

Evidently in the case that $\Omega$ is a ball of radius $R,$ the
critical points that are the  equilibria of $V_{OCP}$  and $V$
coincide as long as we set \ben {3GM \over \Lambda}  = R^3\,, \een
but the values of $V_{OCP}$ and $V$ at the critical points will
differ. In the case that $\Omega$ is not a ball, even the critical
points will differ.

\subsection{ Upper and lower  bounds for the minimum of the energy}
\label{sec:bound}
The following rigorous bounds, whose proofs are discussed in the following
two subsections,  constrain the minimum value of the  energy
\ben
 { 9 \over 10} N (N^{2 \over 3}-1)
\le V^{\rm min } \le
    {9 \over 10} N (N-1 ) ^{ 2 \over 3}  .
\label{lbound}
\een
They are a valuable  check on our numerical results, and it turns out
that the lower bound is a particularly good estimate for the
actual minimum energy. For large numbers of particles our
numerical results support the conjecture that there are many local
minima with energies very close to the lower bound.

\subsubsection{An upper bound}
\label{sec-upper}
The minimum value of a function can never
be greater than the average value of the function over
any sub-domain of its domain. Let us apply this principle
to $V$ which is a function on ${\Bbb R}^{3N}$
and consider
its average value with respect to the uniform distribution
over $(B_3(R_0))^N$
the product of $N$ balls of radius $R_0$, that is, we average
over the sub-domain $0 \le |{\bf r}_a| \le R_0$.
For a pair of particles, and if $n>-3$ the Williamson average
(see Section \ref{sec-Will}) is
\ben
\langle r_{ij}^ n \rangle = {72 (2R_0)^{n} \over (n+3)(n+4)(n+6)}\,,
\een
and thus
\ben
\langle V_{-1} \rangle = { 6 G \over 5 R_0} \sum _{b <a}m_a m_b\,.
\een
On the other hand
\ben
\langle V_2 \rangle = {\Lambda R_0^2\over 10} \sum m_a\,.
\een

\noindent Therefore, the upper bound for the minimum value of $V^{\rm min}$
is, assuming that $\Lambda=
3, G=1, m_a=1$,
\ben
{3 \over 10} NR^2_0 +{ 3\over 5 R_0} N(N-1).
\een
The upper bound will be optimal, that is, smallest,
 if we choose $ R_0^3=(N-1)$.
Substituting back we get

\ben
 V^{\rm min}  \le { 9 \over 10} N(N-1)^{ 2\over 3}.
\een

\subsubsection{ A lower bound for the energy}

As explained in \cite{Bau}, Leib and Narnhoffer \cite{Lieb} proved
a rigorous lower bound for $V_{OCP}$, at least in the case that
$m_1=m_2=\dots =m_N=m$. One defines an ion radius $a$ by 
\ben N {4\pi \over 3} a^3 =A\,. \een
Thus in the case that $\Omega$ is a sphere of radius $R$
\ben
a= {R \over N^{1\over 3}}\,.
\een

Note that in our solutions $a\approx 1$ with considerable
accuracy. Now  one has the extensive lower bound: 
\ben V _{OCP}
\ge -N { 9\over 10} {Gm \over a}\,. \label{lower} \een The
interpretation is that the right hand side of (\ref{lower}) is the
energy of $N$ non-overlapping spheres of radius $a$ with total
charge zero, in other words of $N$ non-overlapping
Thomson type Hydrogen atoms. The packing of these atoms plays an
important role in determining the distribution of the points.

We may re-write the Leib-Narnhoffer  bound (setting $G=m=a=1$) as
\ben V^{\rm min} \ge  { 9 \over 10} N^{5 \over
3} -N{ 9 \over 10}\,. \een

\subsection{Continuum limit}

This has already been alluded to above. It is most
easily obtained by replacing the discrete  distribution of masses
by a continuous density distribution
\ben
\sum m_a \delta ({\bf x}-{\bf r}_a) \longrightarrow \rho({\bf x})\,,
\een
in the variational
problem. Ignoring self-energies, we therefore need to extremize
\ben
{ 1\over 2} G \int \int \rho({\bf x}) \rho({\bf y})
{ 1\over |{\bf x} -{\bf y} |} d^3x d^3y
+ {\Lambda \over 6} \int {\bf x}^2\rho({\bf x}) d^3x + \lambda \int \rho({\bf x}) d^3x\,,
\een
where $\lambda$ is a Lagrange multiplier enforcing the constraint that
the total mass
\ben
M=\int \rho ({\bf x})  d^3x\,,
\een
is fixed. Variation of the density gives a
linear integral equation for $\rho$
\ben
G\int d^3y \rho({\bf y}) { 1\over |{\bf x} -{\bf y} |} + {\Lambda \over 6} {\bf x}^2 +\lambda =0\,.
\een

Acting on this equation with the Laplacian gives
\ben
-4 \pi G \rho +{\Lambda}=0\,.
\een
We have recovered our previous result that the density must be constant.
But it is clear that the density cannot be everywhere
constant and  still satisfy the constraint that the total mass be
fixed.
Moreover we have not deduced that the boundary of the
blob of uniform fluid must be
spherically
symmetric. This is presumably because we have not been sufficiently
careful
about boundary  effects in the variation.

\subsection{Separation probability distribution}
\label{sec-Will}

The cumulative  probability for the separation of two points ${\bf
r}_1$ and ${\bf r}_2$ uniformly distributed inside a sphere of radius
$R$ seems to have been given originally by Williamson using an
extremely ingenious geometrical argument~\cite{Williamson}. 
Below we rederive this result using differential forms.

The volume form on ${\Bbb R}^3 \times {\Bbb R} ^3 $ is
given in spherical polars about some fixed axis with origin $O$ by
\ben
\omega =r_1 ^2 \sin \theta_1 dr_1 \wedge d\theta_1  \wedge d \phi_1 \wedge
  r_2^2 \sin \theta_2 dr_2 \wedge d \phi_2\,.
\een

Consider the triangle $O12$ with sides of length $r_1, r_2 $ and $r_{12}$.
Let $\psi$ be the angle $O12$ and $\chi$ the angle of the plane of
 the triangle about an  axis along the side $O1$.
Then by means of a rotation of the second set of spherical
polars one has
\ben
\omega =r_1^2 \sin \theta_1 dr_1 \wedge d \theta _1 \wedge
d \phi_1 \wedge  r_{12}^2 \sin \psi
 dr_{12} \wedge d \psi \wedge d \chi\,.
\een

Now the cosine formula for the triangle tells us that
\ben
r_2^2 =r_1^2 + r_{12}^2 -2 r_1 r_{12} \cos \psi\,,
\een
and therefore 
\ben
r_2 d r_2 = (r_1 -r_{12} \cos \psi )dr_1 + (r_{12}
-r_1 \cos \psi) d r_{12}+ r_1 r_{12} \sin \psi d \psi\,.
\een
Eliminating $d\psi$ gives
\ben
\omega = r_1 r_2 r_{12}  \sin \theta _1 dr_1   \wedge dr_2 \wedge
 dr_{12}
 \wedge d \theta_1 \wedge d \phi _1 \wedge d \chi.
\een
The integrals over $\theta_1 , \phi_1 , \chi$ may be done immediately
so that
\ben
\omega= 8 \pi^2 r_1 r_2 r_{12}  dr_1 \wedge dr_2 \wedge dr_{12}.
\een

In order to obtain $dP$ we set $r=r_{12}$ and
integrate over $r_1$ and $r_2$
consistent with the points $1$ and $2$ being confined to lie inside
a ball of radius $R$ and divide by $16 \pi^2 R^6/9$.
To perform the integration it is convenient to introduce
the coordinates
$x=r_1 +r_2$ and $y= r_1-r_2$.
The ranges of integration are obtained by applying the triangle inequalities
and are given by $r \le x \le 2R -|y|$ and $|y|\ge r$.

The result is 
\ben
{\rm Prob} (|{\bf r}_1-{\bf r}_2 | \le  r)= {r^3 \over R^3 }-{ 9r^4
\over 16 R^4} + {r^6 \over 32 R^6}\,.
\een
One has of course $ {\rm Prob} (|{\bf r}_1-{\bf r}_2 | \le 2R)=1$.
The probability density is thus
\ben
dP=p(r)dr= \Bigl ({3r^2 \over R^3} - {9 r^3 \over 4 R^4} +
{3r^5 \over 16 R^6}  \Bigr )dr\,,
\label{Willdist}
\een
from which, the mean separation is
\ben
{\langle  r \rangle }=\int _0^{2R} rp(r)dr = { 36 \over 35} R
\approx 1.02857R\,.
\label{Willmean}
\een

The  numerical results described later agree with this rather well.
In what follows we shall denote averages with respect to
the Williamson distribution as above and averages taken over our
numerically generated set of points (or pairs of points in this case)
by an overbar. Thus numerically, as we shall show, we find that
${\bar r}\approx \langle r \rangle$
to a good accuracy.
Of course to compare we must say what the value of $R$ is.
This will usually be done using the formula $R=(N-1)^{ 1 \over 3}$.
Recall that this relation between $R$ and $N$ is the one derived
in Section \ref{sec-upper} in order to make the upper bound on $V^{\rm min}$
optimal. 

The most likely separation is given by the  root
between $r=0$ and $r=2R$
of the cubic equation
\ben
6- {27 r \over 4R} + {15 r^ 3 \over 16 R^3}=0\,.
\een
Because $r=2R$, is a root, the cubic factorizes
\ben
( r-2R) ( 5r^2+10Rr-16R^2)=0\,,
\een
and the solution we want is
\ben
r= {R\over 5}(\sqrt{105}-5 ) \approx 1.04939R\,.
\een

\subsection{Distribution of triangles}

Later we shall present the statistics of triples of points
computed numerically. For points uniformly distributed inside
a sphere an interesting quantity to consider is the distribution
of angles over all triangles given by any three points.
Unfortunately, it appears that no analytic
expression for this distribution of angles is known.
Deriving a formula for this
distribution would therefore seem to be a very worthwhile
exercise in geometric probability.
One result that is known is that the probability that any
angle is acute is given by $33/70$~\cite{Williamson,TRIAN2}. Numerically we shall
find a good agreement with this value.

\subsection {Point charges on a sphere}
\label{sec-sphere}
The problem here is to minimize the potential energy $V_{-1}$ subject to the
constraint that the points lie on a sphere of some given radius.
For reasons which are unclear to us, this problem has come to be
associated with Thomson's name even though he appears not to have
posed it explicitly. What he had in mind is perhaps that given
the existence of a shell structure then one only needs to minimize
the energy with respect to positions inside the shell. A large number
 of papers have investigated this problem; see \cite{sphere} and 
references therein for details.

To see this more explicitly,  note that in order  to enforce the
constraint one introduces $N$ Lagrange multipliers $\Lambda_a$.
One obtains the equations

\ben
{ \Lambda_a \over 3} m_a{\bf r}_a + \sum_{b\ne a} {Gm_am_b
({\bf r}_b -{\bf r}_a) \over |{\bf r}  _a -{\bf r}_b |^3 }=0. \label{sphere}
\een

The interpretation of the first term in (\ref{sphere}), $ {\bf
F}_a={ \Lambda_a \over 3} m_a{\bf r}_a $ is that it is the inward
force exerted on the particle  necessary to counteract the outward
repulsion of the remaining particles. Thus on
solving the equations and constraints, the Lagrange multipliers
$\Lambda_a$ will turn out to be positive. If it happens that
all the $\Lambda_a$'s are equal then this is also a 
central configuration (it is a solution of (\ref{defn1})). This may
be true only approximately if the distribution of points is
sufficiently spherically symmetric. 

\subsection{Sphere packing problems}

As we have indicated above, there appears to be
 a close relation
between central configurations and the classical sphere packing
problem: to prove that there is no
packing of congruent spheres in three dimensions  with density or
packing ratio $\eta$ exceeding that of a face-centred cubic (FCC) with
$\eta=\pi/\sqrt{18} \approx 0.74048$. This long-standing conjecture, due
originally to Harriot and Kepler has now been proved by Hales (see
ref.~\cite{Hales} for an overview and references). 

The highest
  packing density  is achieved for FCC packing which is
crystallographic, but it is well-known that there are uncountably many
other packings, both crystallographic and non-crystallographic with
the same packing density. Thus, viewed as an optimization problem, the
sphere packing problem has infinitely many optima with essentially the same
density. Moreover local optima with vacancies, that is with  a finite
number  of  isolated spheres missing, have in the infinite limit the
same density.  In the case of finite sphere packings there will
clearly be many local optima very close to the closest packing. This
feature is certainly shared by central configurations.  
            
The comparison of central configurations with sphere packings can be
taken further. For example, a key fact about any sphere packing is,
as stated first in print by Halley~\cite{Halley} in connection with
his prior account of Olber's Paradox,  that at most 12 congruent
spheres may touch a thirteenth congruent sphere.  In other words, the
maximum coordination number (that is  the number of nearest
neighbours) for  close-packing is 12. This fact, asserted by Newton
and denied by Gregory \cite{Greg}, 
would be a useful diagnostic tool in assessing whether our configurations
are close-packed (they are certainly not FCC) but unfortunately for
central configurations there is no unambiguous way to define a 
coordination number, and any numerical results computed are very
sensitive to its definition.

One may
refine the above discussion a little \cite{Max}.  The local cell for FCC
packing is a rhombic dodecahedron. However, the local cell of smallest
volume is a regular pentagonal dodecahedron.  This cannot, because of
it's five-fold symmetry, give a lattice packing of course but it can
appear in small clusters and this happens in our case for 13
particles. In the same note it is  remarked that most physicist's
believe that the optimum for the One Component Plasma is a
body-centred cubic (BCC) packing. As we discuss in Section \ref{sec-cry}
we have seen little evidence for
that in our results. It is perhaps worth remarking here that the
published energies of various lattices in the One Component Plasma
problem~\cite{foldy}
seem to be extremely close and this alone indicates it shares
with the sphere packing problem the feature that there are many
critical points very close to the minimum.
It turns out to be  worth exploring in more detail some further
features of sphere packings since they have some diagnostic value
 in understanding our numerical results. This is especially true in
connection with the shell structure which will be discussed in Section 4.

\section{Case I : small numbers of points}\news

Small numbers may be studied analytically and numerically;
historical information may be found for example in ref.~\cite{Win} or
ref.~\cite{Hag}, and we largely ignore planar
solutions since (for $N>3$) these appear  to be unstable.
By symmetry, one expects any regular polyhedron to provide
a solution but not necessarily  a stable one.
One can also place a mass point at the centre
of a regular polyhedron.
For the same reason
it is also clear that pyramidal  and bi-pyramidal solutions
should exist for arbitrary numbers of particles as well as prism
and anti-prism solutions. Again, placing a mass point at the centre
of bi-pyramids, prisms and anti-prisms is possible.
According to Hagihara \cite{Hag},  Blimovitch \cite{Blim1,Blim2} claims
two similar and similarly situated regular polyhedra are possible,
as well as a regular polyhedron together with its dual.

\subsection{N=3 Lagrange's triangle}

Relative equilibria are planar solutions of (\ref{defn1})
and include collinear solutions. They may also give rise to
 rigidly rotating solutions of Newton's equations of motion.
Planar configurations will be the subject of another paper
and so here we will restrict attention to the
case when $N=3$.
In that case, for arbitrary masses,
 the only non-collinear solution
is Lagrange's  equilateral triangle.
In standard units the sides of the triangle are $^3\surd 3 = 1.4422496\dots$
which is larger than the distance $^3 \surd 2=1.259921\dots$ of the dipole.
In what follows it will be useful to envisage Lagrange's
solution  as three spheres
touching one another.
For some interesting recent work on the planar
case including the relation to a hard disc model and with applications
to the final shapes of systems of particles moving under repulsive
inverse square law
forces, see refs.~\cite{Glass1,Glass2}. For other work on planar configurations see ref.~\cite{klemper}. If one really were dealing with two
dimensions, then the analogous problem would involve
a logarithmic potential; for results on this case see ref.~\cite{Kog}.

\subsection{N=4 : tetrahedral configuration}

The first non-planar case is for $N=4$. The existence of a regular
tetrahedral solution for arbitrary  positive values of
$(m_1,m_2,m_3,m_4)$ was shown by Lehmann Filh\'es in 1891
\cite{Fil} and the uniqueness among all non-planar solutions by
Pizetti in 1903 \cite{Piz}.

The existence is obvious by noting that if we choose side
length $({3GM
\Lambda} )^{ 1\over 3}$ for our tetrahedron then by (\ref{distance})
{\sl every } two body force will vanish. The necessity follows by
noting that if the four are not co-planar, then the six
inter-particle distances $r_{ab}$, $1\le a <b \le 4$ give six
independent coordinates on $C_4({\Bbb R} ^3 )/E(3)$ and so the
potential function must be stationary with respect to independent
variations of all six inter-particle distances. From
(\ref{manybody1}) and  (\ref{manybody2}), it follows that every
inter-particle distance must be a stationary point of the function
$U$ in (\ref{manybody2}).

 In normalized
units the side of the tetrahedral configuration, which should be
envisaged as four mutually touching close packed spheres is $^3
\surd 4 =1.5874011\dots$
The significance of the tetrahedron as far as our work is
concerned is that it not infrequently
seems to occur as a sub-configuration
inside a nested set of shells.

\subsection{N=5 : triangular bi-pyramid}
\label{sec-N=5}

Surprisingly this is not completely understood \cite{Sch, Fay}.
Numerically one finds a minimum in the form of triangular
bi-pyramid. In addition one knows that there is a solution with one point
at the centre of a tetrahedron and a pyramidal solution on a
square base \cite{Sch, Fay}. It is not difficult to imagine other,
presumably unstable, solutions.

The bi-pyramid is not regular. However, it closely resembles 
a bi-pyramidal cluster  obtained by close-packing 5 equal spheres.
The three points which form the equilateral triangle are at a
distance of 1.081 from the origin, whereas the two remaining points
are at a distance of 1.104 from the origin. 
In terms of edges lengths we can summarize this information in 
Table~\ref{el5}. For each type of vertex we give its multiplicity
(the number of times such a vertex occurs in the configuration),
its valency (the number of nearest neighbours), and the edge lengths
of the polyhedron given by the distances of the nearest neighbours.
The numbers in round brackets after each edge length denote the
multiplicity of this nearest neighbour length.
Note that each edge of the polyhedron is represented twice, since
we deal with each vertex individually.
\begin{table}\centering\begin{tabular}{|c|c|c|}\hline
Multiplicity&Valency&Edge Lengths\\ \hline
3&4&1.545(2),1.872(2)\\
2&3&1.545(3)\\
\hline\end{tabular}
\caption{For the $N=5$ polyhedron we list the multiplicity of each type
of vertex, its valency, and its edge lengths together with their
multiplicities (given as the number in brackets after each edge length).
}
\label{el5}\end{table}

The information in Table~\ref{el5} therefore summarizes the fact that
there are three 4-valent vertices (the ones which form the equilateral
triangle) and two 3-valent vertices (the ones which sit above and below
the equilateral triangle). The equilateral triangle has edge length
$1.872$ but the six remaining edge lengths are all shorter at $1.545.$
Taking the average of the nine edges lengths gives $\bar l=1.654,$
which is in good agreement with the diameter $d=1.65$ which we 
use in our sphere packing model.

It is interesting to note that the same triangular bi-pyramid
also arises as the energy minimizing configuration using a scale
invariant energy function \cite{AS} and the ratio of the two distances
from the origin  $1.081/1.104=0.979$ is precisely the same value
as obtained in that case. 
In fact for all $N\le 12$ the configurations
of minimizing points appear to be remarkably similar
for the two problems (taking into account the 
scale invariance of one of the energy functions). 


\subsection{$6 \le N \le 12$}

In this range the minima form a single shell.
If $N=6$ we have an
octahedron, with edge length $1.676.$ 
If $N=7$ we have a pentangular bi-pyramid. The five points forming
the pentagon sit on a circle of radius $1.283$ and the remaining
two points are at a distance of $1.248$ from the origin. The
ratio of these two distances $1.283/1.248=1.028$ is again
equal to that for the pentangular bi-pyramid which results
from minimizing the scale invariant energy function of ref.~\cite{AS}.
In terms of edge lengths this information is summarized in Table~\ref{el7}.
The average edge length is $\bar l=1.696.$
\begin{table}\centering\begin{tabular}{|c|c|c|}\hline
Multiplicity&Valency&Edge Lengths\\ \hline
5&4&1.509(2),1.790(2)\\
2&5&1.790(5)\\
\hline\end{tabular}
\caption{Vertex types and edge lengths for the $N=7$ polyhedron.}
\label{el7}\end{table}

$N=8$ is the first example in which some of the
faces are not triangular, it being a square anti-prism, obtained from a cube
by rotating the top face by $45^\circ$ relative to the bottom face. 
Each vertex is 4-valent and contains two edges of length
$1.581$ and two of length $1.738,$ giving an average length $\bar l=1.660.$

\begin{table}\centering\begin{tabular}{|c|c|c|}\hline
Multiplicity&Valency&Edge Lengths\\ \hline
3&4&1.615(4)\\
6&5&1.615(2),1.742(2),1.989(1)\\
\hline\end{tabular}
\caption{Vertex types and edge lengths for the $N=9$ polyhedron.}
\label{el9}\end{table}
For $N=9$ the
points lie on the vertices of three parallel equilateral triangles, with the
middle triangle rotated by $60^\circ$ relative to the other two.
The edge lengths are given in Table~\ref{el9} and the average is 
$\bar l=1.705.$

\begin{table}\centering\begin{tabular}{|c|c|c|}\hline
Multiplicity&Valency&Edge Lengths\\ \hline
8&5&1.600(1),1.621(2),1.898(2)\\
2&4&1.600(4)\\
\hline\end{tabular}
\caption{Vertex types and edge lengths for the $N=10$ polyhedron.}
\label{el10}\end{table}
The $N=10$ polyhedron can be obtained from the $N=8$ one by replacing each
square by a hat made from four triangles with a 4-valent vertex.
The edge lengths are given in Table~\ref{el10} and the average is 
$\bar l=1.706.$

\begin{table}\centering\begin{tabular}{|c|c|c|}\hline
Multiplicity&Valency&Edge Lengths\\ \hline
1&6&1.524(2),1.977(4)\\
2&5&1.624(2),1.730(2),1.825(1)\\
2&5&1.624(2),1.659(1),1.806(2)\\
4&4&1.511(1),1.546(1),1.730(1),1.806(1)\\
2&4&1.524(1),1.546(2),1.659(1)\\
\hline\end{tabular}
\caption{Vertex types and edge lengths for the $N=11$ polyhedron.}
\label{el11}\end{table}

For $N=11$ the polyhedron contains a vertex with six nearest neighbours.
The existence of the
single vertex with six neighbours means that this configuration
is not very symmetric.
The edge lengths are given in Table~\ref{el11} and the average is 
$\bar l=1.680.$

$N=12$ forms
a regular icosahedron with edge length $\bar l=1.682.$

We have already commented that these configurations occur as
the minima of a scale invariant energy function and furthermore,
as discussed in that situation \cite{AS}, the associated polyhedra
are of the same combinatoric type as those associated with the
solution of the points on a 
sphere problem discussed in Section \ref{sec-sphere}. 
In fact,
the correspondence is more than a combinatoric match since a projection
of the points onto the sphere appears to produce the solutions of
the sphere problem.

\begin{figure} 
\begin{center}
\leavevmode
\ \vskip -0cm
\epsfxsize=15cm\epsffile{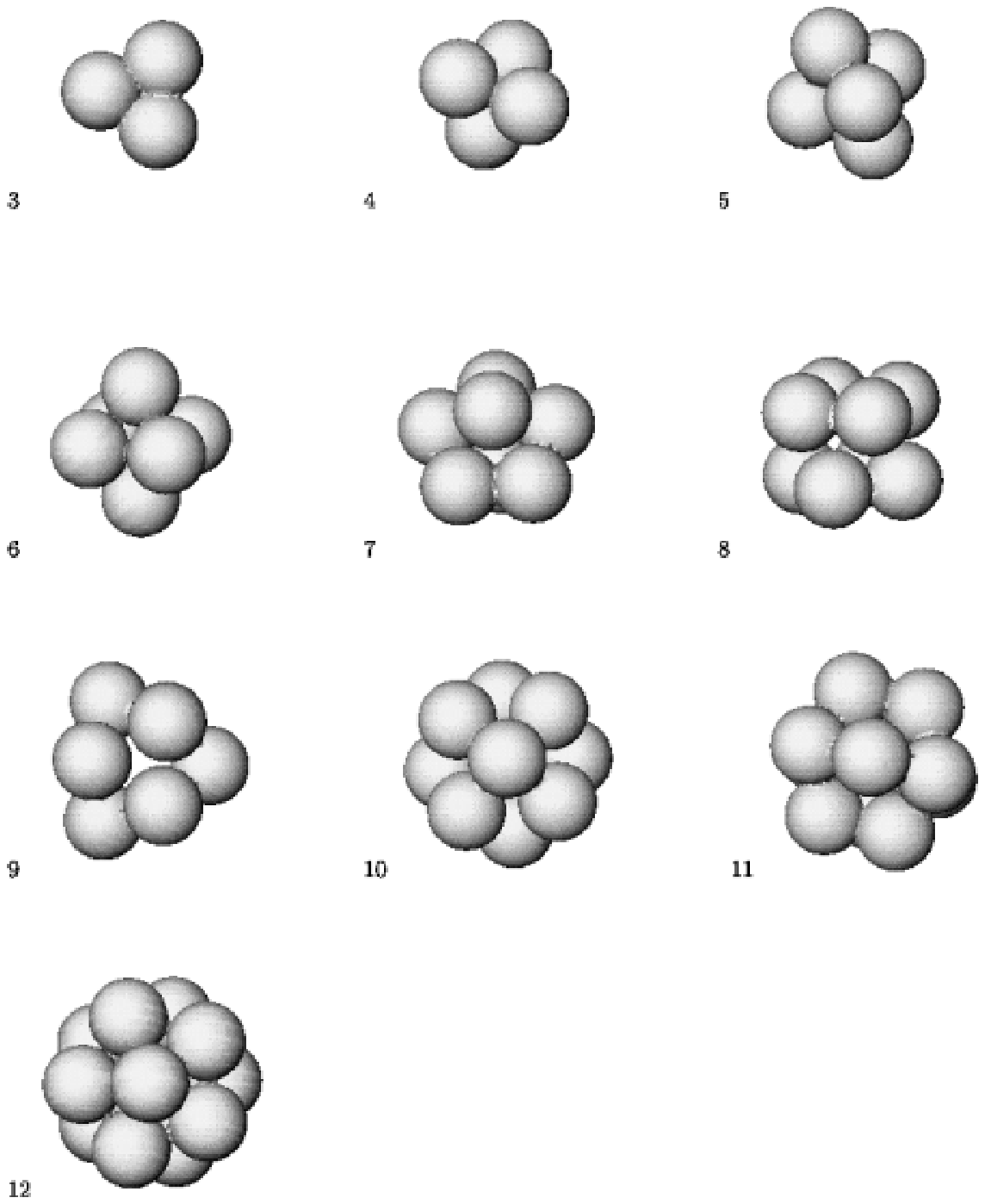}
\caption{For  $3\le N \le 12$ we display our configurations of $N$ points
by plotting spheres of diameter $d=1.65$ around each of the points.
\goodbreak {\em See 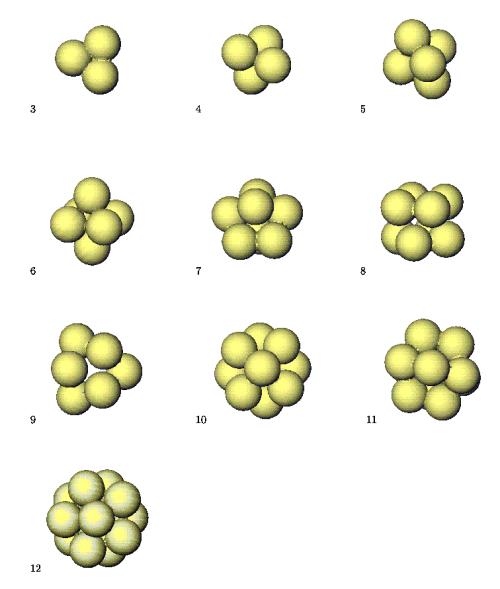 for a colour version of this figure.}}
\label{balls}
\end{center}
\end{figure}

In fig.~\ref{balls} we display our configurations of points,
for $3\le N \le 12$, by plotting spheres of diameter $d=1.65$
around each of the $N$ points. This highlights the similarity to
sphere packing configurations.

\subsection{Deltahedra}

A  regular deltahedron is a polyhedron  all of whose faces
are equilateral triangles. A combinatoric deltahedron
is a polyhedron of the same combinatoric type.
For any deltahedron
we have $2E=3F$
(in the following $E,F,V$ refer to the number of
edges, faces and vertices of a polyhedron).
 If it has the topology of a sphere we have
$F-E+V=2$, and thus
\ben
F=2(V-2), \qquad E=3V-2.
\een
There are just 8 convex regular deltahedra. They have
$V=4,5,6,7,8,9,10,12$.  There is no convex deltahedron with $V=11$.

The minimum energy solutions for
$N=4,5,6,7,9,10,12$ closely resemble (or are) regular deltahedra,
as can be seen from the tables of edge lengths. In each of
these cases there are no more than three different edge
lengths forming the polyhedron and they are all reasonably
close in value.
This is related
to the geometry of deltahedra
 taken together with the existence of Lagrange's
triangular solution
and the fact that
a particular spacing is picked out at which the inter-particle
force vanishes.

The forces on regular convex deltahedra
are almost in equilibrium and presumably
only require
small adjustments to cancel exactly. 
 The fact that the $N=11$ configuration is not regular
is automatic, since, as mentioned above,
 no regular deltahedron exists with 11 vertices.
From Table~\ref{el11} it can be seen that there are
many different edge lengths forming the $N=11$ polyhedron.

\begin{figure} 
\begin{center}
\leavevmode
\epsfxsize=15cm\epsffile{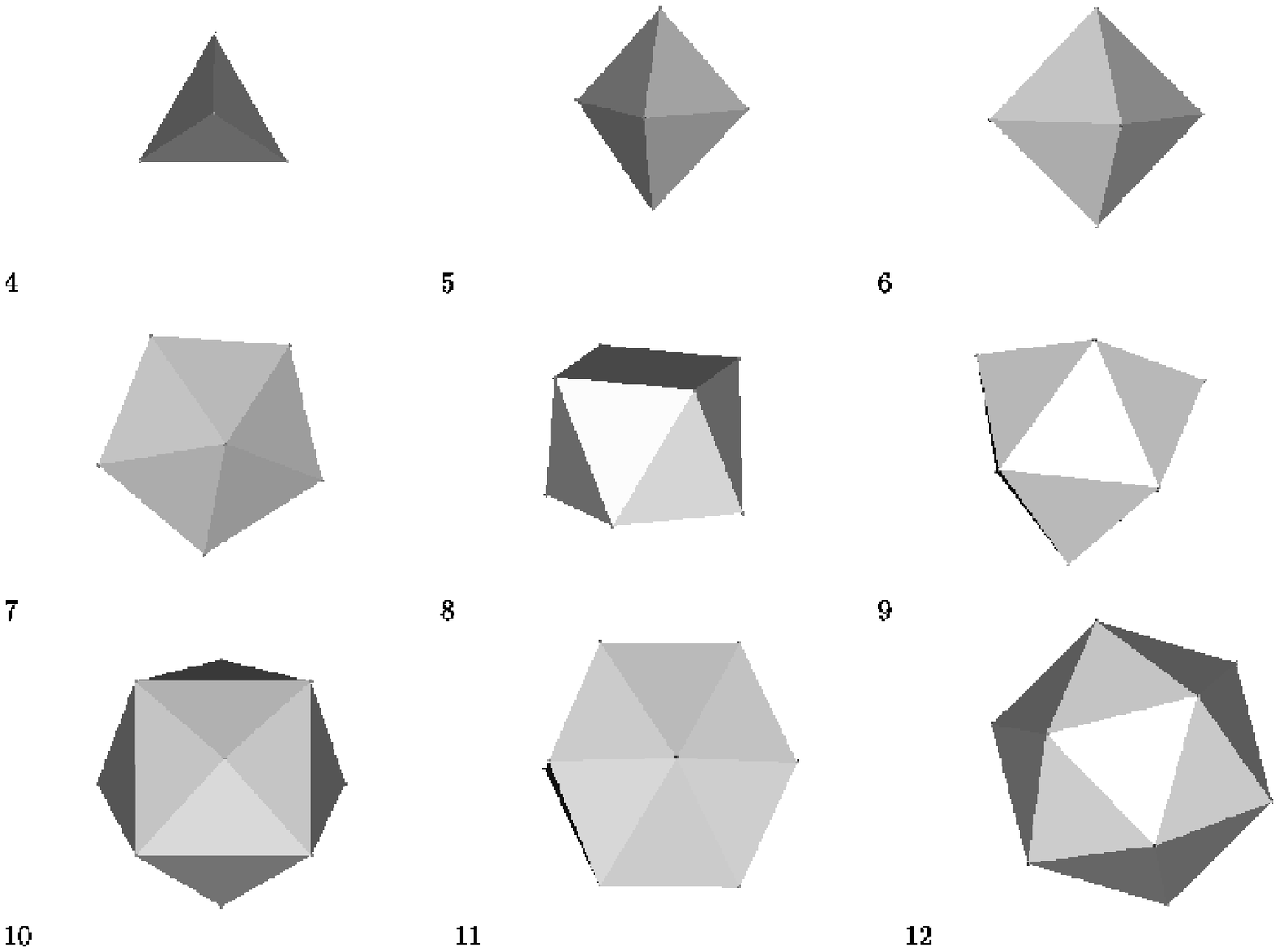}
\vskip -0cm
\caption{The polyhedra associated with the set of $N$ points for $4\le N \le 12.$}
\label{poly}
\end{center}
\end{figure}

The actual polyhedra themselves are displayed in fig.~\ref{poly}
for $4\le N \le 12.$

\small
\begin{table}
\ \vskip -3.2cm
\centering\begin{tabular}{|r|r|r|r|||r|r|r|r|}\hline
$N$ & $V$\quad\quad\ & $V/\mbox{Bound}$ & Shells\ \
& $N$ & $V$\quad\quad\quad & $V/\mbox{Bound}$ & Shells\ \ \\ \hline
   1&          0.0000&  1.000000&01/00/00&  51&      586.1361&    1.001364&10/41/00\\
   2&          1.1906&  1.126006&02/00/00&  52&      606.0110&    1.001339&10/42/00\\
   3&          3.1201&  1.069919&03/00/00&  53&      626.1669&    1.001335&10/43/00\\
   4&          5.6696&  1.036227&04/00/00&  54&      646.5716&    1.001305&10/44/00\\
   5&          8.9100&  1.029100&05/00/00&  55&      667.2312&    1.001261&12/43/00\\
   6&         12.6391&  1.016787&06/00/00&  56&      688.1384&    1.001197&12/44/00\\
   7&         17.0243&  1.016157&07/00/00&  57&      709.3484&    1.001194&12/45/00\\
   8&         21.8643&  1.012236&08/00/00&  58&      730.8185&    1.001192&01/12/45\\
   9&         27.2144&  1.009937&09/00/00&  59&      752.5386&    1.001178&01/12/46\\
  10&         33.0575&  1.008642&10/00/00&  60&      774.5108&    1.001159&12/48/00\\
  11&         39.4041&  1.008647&11/00/00&  61&      796.7202&    1.001117&01/12/48\\
  12&         46.0883&  1.006118&12/00/00&  62&      819.2150&    1.001116&01/13/48\\
  13&         53.3116&  1.006132&01/12/00&  63&      841.9375&    1.001085&01/14/48\\
  14&         60.9584&  1.006069&01/13/00&  64&      864.9324&    1.001079&01/14/49\\
  15&         68.9578&  1.005074&01/14/00&  65&      888.1564&    1.001050&01/14/50\\
  16&         77.3816&  1.004509&01/15/00&  66&      911.6371&    1.001031&01/15/50\\
  17&         86.2009&  1.004020&01/16/00&  67&      935.3691&    1.001017&01/15/51\\
  18&         95.4178&  1.003699&01/17/00&  68&      959.3390&    1.000994&01/16/51\\
  19&        105.0215&  1.003470&01/18/00&  69&      983.5543&    1.000973&01/16/52\\
  20&        115.0418&  1.003635&01/19/00&  70&     1008.0264&    1.000965&01/16/53\\
  21&        125.3808&  1.003364&01/20/00&  71&     1032.7387&    1.000955&01/16/54\\
  22&        136.1199&  1.003403&01/21/00&  72&     1057.6937&    1.000947&01/17/54\\
  23&        147.2015&  1.003330&02/21/00&  73&     1082.8858&    1.000935&01/17/55\\
  24&        158.6157&  1.003140&02/22/00&  74&     1108.3176&    1.000925&01/17/56\\
  25&        170.4147&  1.003193&02/23/00&  75&     1133.9875&    1.000914&01/18/56\\
  26&        182.5115&  1.002991&02/24/00&  76&     1159.9000&    1.000909&01/18/57\\
  27&        194.9551&  1.002855&03/24/00&  77&     1186.0483&    1.000903&01/18/58\\
  28&        207.7545&  1.002841&03/25/00&  78&     1212.4297&    1.000896&01/18/59\\
  29&        220.8612&  1.002723&04/25/00&  79&     1239.0530&    1.000896&01/18/60\\
  30&        234.2757&  1.002540&04/26/00&  80&     1265.9012&    1.000888&01/20/59\\
  31&        248.0035&  1.002339&04/27/00&  81&     1292.9691&    1.000872&01/20/60\\
  32&        262.0781&  1.002265&04/28/00&  82&     1320.2933&    1.000875&02/20/60\\
  33&        276.4994&  1.002311&04/29/00&  83&     1347.8394&    1.000871&02/21/60\\
  34&        291.1997&  1.002238&04/30/00&  84&     1375.6035&    1.000859&02/21/61\\
  35&        306.2062&  1.002160&05/30/00&  85&     1403.5980&    1.000850&02/21/62\\
  36&        321.5036&  1.002043&06/30/00&  86&     1431.8213&    1.000841&02/21/63\\
  37&        337.0954&  1.001909&06/31/00&  87&     1460.2694&    1.000831&02/22/63\\
  38&        352.9683&  1.001731&06/32/00&  88&     1488.9476&    1.000825&02/22/64\\
  39&        369.2331&  1.001823&06/33/00&  89&     1517.8678&    1.000830&03/22/64\\
  40&        385.7436&  1.001779&06/34/00&  90&     1546.9950&    1.000824&03/22/65\\
  41&        402.5671&  1.001788&06/35/00&  91&     1576.3474&    1.000819&03/23/65\\
  42&        419.6643&  1.001757&07/35/00&  92&     1605.9053&    1.000803&03/22/67\\
  43&        437.0420&  1.001710&07/36/00&  93&     1635.6839&    1.000788&03/24/66\\
  44&        454.6979&  1.001650&08/36/00&  94&     1665.6841&    1.000774&03/24/67\\
  45&        472.6332&  1.001585&08/37/00&  95&     1695.9225&    1.000770&04/24/67\\
  46&        490.8654&  1.001556&08/38/00&  96&     1726.3794&    1.000767&04/24/68\\
  47&        509.3649&  1.001505&09/38/00&  97&     1757.0511&    1.000761&04/25/68\\
  48&        528.1383&  1.001451&09/39/00&  98&     1787.9348&    1.000753&04/25/69\\
  49&        547.1978&  1.001420&09/40/00&  99&     1819.0277&    1.000740&04/26/69\\
  50&        566.5327&  1.001394&09/41/00& 100&     1850.3349&    1.000727&04/26/70\\
\hline\end{tabular}
\caption{For $N\le 100$ we list the minimum
energy $V,$ the ratio of this energy to the lower bound, and the
number of points in each shell. }
\label{1to100}\end{table}

\normalsize

\subsection{$N\le 100$}
In Table~\ref{1to100}
 we present the minimal value of the energy for all $N\le 100.$
The ratio of this energy to the value of the lower bound (\ref{lbound}) is
also given, from which it is clear that the lower bound is an extremely
tight one and that the percentage excess over this bound decreases with
increasing $N.$ These results were obtained using both a multi-start simulated 
annealing algorithm and a multi-start gradient flow code. Both methods were
applied independently and led to the same common results. We therefore believe
that the configurations we have found are the global minima for each value
of $N$ and the energies are accurate to the level quoted. As expected, during
our computations we found large numbers of local minima, which we shall ignore.
The same codes were used to generate the configurations discussed later in the
paper with $N\gg 100,$ but we make no claim that they are the global minima, merely
that they are local minima (which may or may not be global) whose
energies we expect to be very close to that of the global minimum.

For $N\le 12$ all the points lie close to the surface of a sphere, but this
is not the case for $N>12.$ In particular for $N=13$ there are 12 points
on the vertices of a regular icosahedron and an additional single point
at the origin. We denote this structure by the code $01/12,$ indicating
that there are two shells, the first one containing a single point
and the second containing 12 points. For $N\le 100$ there are at most
three shells. In Table~\ref{1to100} we present the shell structure for the
minimal energy configuration by listing its code as above. We find that
within each shell the arrangement of points resembles that for the solution
of the problem of Section 2.8.
 For example, if an inner shell contains four points
then they are located on the vertices of a regular tetrahedron.

\begin{figure} 
\begin{center}
\leavevmode
\epsfxsize=12cm\epsffile{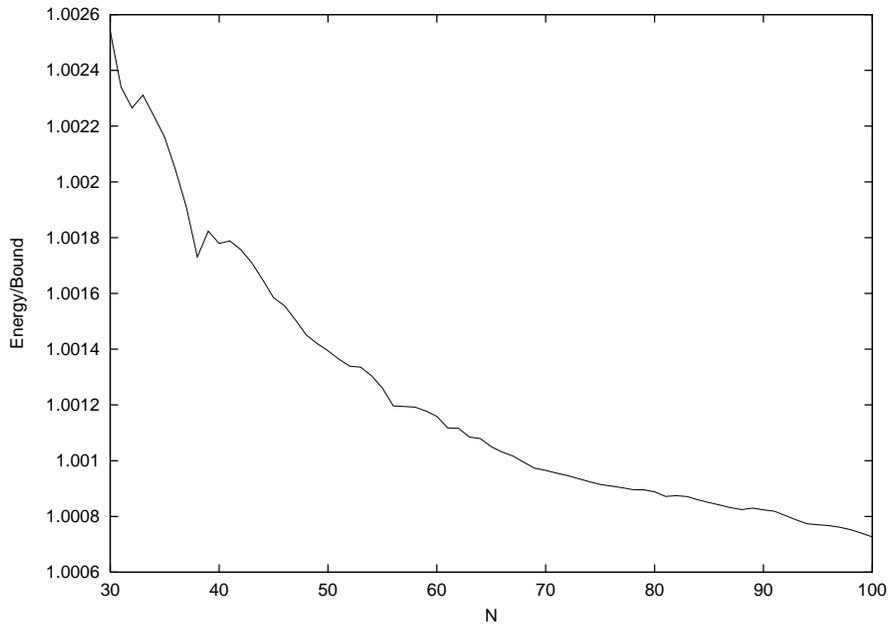}
\caption{The ratio of the energy to the bound for $30\le N\le 100.$}
\label{magic}
\end{center}
\end{figure} 

\begin{figure} 
\begin{center}
\leavevmode
\epsfxsize=12cm\epsffile{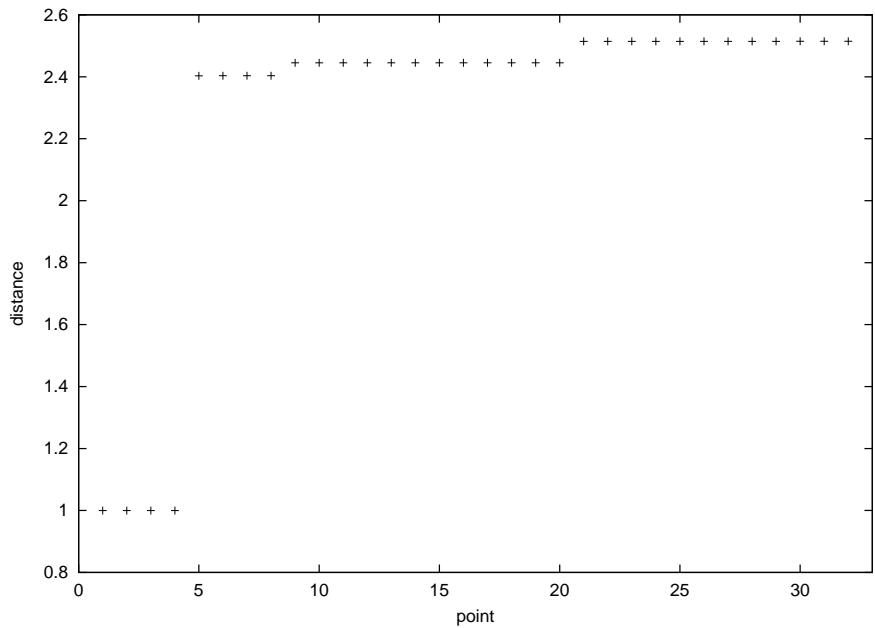}
\caption{The distance of each point from the origin
for the $N=32$ minimal energy configuration.}
\label{32}
\end{center}
\end{figure}

In fig.~\ref{magic} we plot the ratio of the energy to the bound for
$30\le N\le 100.$ From this plot we see that there are magic numbers
at which this ratio drops more sharply than usual. The most striking
examples are $N=32$ and $N=38.$ These magic numbers occur when two
(or more) shells both have a large symmetry group. For $N=32$ we see
from Table~\ref{1to100}
 that the shell structure is $4/28$, that is, there are two
shells with the inner shell containing 4 points and the outer shell
containing 28 points. The solution of the sphere problem for 4
points is a regular tetrahedron, while the solution of the sphere
problem for 28 points also has tetrahedral symmetry.
These two solutions, if appropriately aligned as inner and outer shells 
can therefore preserve
tetrahedral symmetry, and this is precisely the arrangement we 
find for the $N=28$ configuration. In fig.~\ref{32} we plot the distance
of each of the 32 points from the origin. We see that within the
second shell there is a substructure consisting of three mini-shells,
each of which contains a multiple of four points, consistent with
the tetrahedral symmetry. A similar situation arises for $N=38$,
with the shell structure being $6/32.$ The solution of the sphere
problem for 32 points has icosahedral symmetry, the associated polyhedron being
the dual of the truncated icosahedron, while for 6 points the solution
of the sphere problem is an octahedron. The $N=38$ solution consists
of these two nested solutions aligned to preserve their common tetrahedral
subgroups.

\section{Case II : moderate numbers of points}\news

We have discussed in the previous Section that there is a transition
from a single shell when $N=12$ to two when $N=13$, and that a similar
transition, from two to three shells, takes place on passing from
$N=57$ to $N=58$. One might expect that further transitions take place
as one increases $N$ and indeed this is the case, although as we will
discuss in the subsequent Section this eventually gives way to what is
effectively a uniform distribution for $N>1000$. Here,
we will discuss and quantify the structure of the shells.

\subsection{Shell structure}

Considerations based on Newton's theorem make it plausible that, at
least for moderately large numbers of points, one may expect solutions
made up of nested shells, each one giving an approximate solution 
to the problem of Section 2.8 
and arranged in such a way that the average density
of points is uniform. More precisely if $M_k$ is the total mass within
the $k$'th  shell which has radius $R_k$ then
\ben
3GM_k=\Lambda R^3_k \label{equipoise}.
\een
In what follows it will be convenient to count the shells
from the centre so that if there are $S$ shells, the
last shell has radius $R_S$ and $M_S=M$.
The mass of the $k$'th shell is
\ben
\Delta M_k= M_k-M_{k-1},
\een
and thus 
\ben
\Delta M_k = { \Lambda \over 3G}\left( R_k^3-R_{k-1}^3\right)\label{jump}.
\een
We can analogously define
\ben
\Delta  N_k = N_k-N_{k-1},
\een
where $N_k$ is the number of particles inside and on the $k$'th shell. For each shell we can define the surface density to be $\sigma_k=\Delta N_k/(4\pi R_k^2)$.

Our numerical calculations suggest that, ignoring a single particle or pairs of
 particles which might congregate at the centre, the radii of the shells are 
in arithmetic progression and that the surface density of particles in each shell
is approximately constant;
 the relevant constants being almost universal between different configurations.
 Let us define the constants $R_{\rm c}$ and $R_0$ such that
 $R_k\approx kR_{\rm c}-R_0$ and set $\sigma_k\approx\sigma.$
Then  $\Delta N_{k}\approx 4\pi\sigma R_k^2$ and hence for large $k$ we see that 
$\Delta N_k\propto k^2$. This provides an interesting approximation which appears
 to have some veracity if one ignores the innermost shells.
 Moreover, one can use the virialization condition (\ref{virial}) to give 
\ben
V\approx{3\over 2}\sum_{i=1}^{S}R_i^2\Delta N_i\approx 6\pi\sigma\sum_{i=1}^{S}R_i^4
\approx 6\pi\sigma \sum_{i=1}^S (iR_{\rm c}-R_0)^4\,.
\label{roughv}
\een
This estimate for the energy, while nowhere near as accurate as 
that discussed
 in Section \ref{sec:bound}, is accurate to within a few percent.

\subsection{Worked example}

\begin{table}\centering\begin{tabular}{|c|c|c|c|c|}\hline
$k$&$\Delta N_k$&$R_k$&$\delta R_k$&$4\pi\sigma_k$\\ \hline
1&6&1.227& 0.029 &3.983\\
2&31&2.655& 0.150 &4.399\\
3&78&4.148& 0.150 &4.533\\
4&149&5.678& 0.129&4.622\\
5&236&7.197& 0.037&4.556\\
\hline\end{tabular}
\caption{The shell structure for a configuration with $N=500$ points.
 Note that the radii of the shells are approximately in arithmetic
 progression and that the value of $\sigma_k$ is roughly constant.}
\label{example}\end{table}

\begin{figure} 
\begin{center}
\leavevmode
\vskip -0cm
\centerline{\epsfxsize=15cm\epsfxsize=15cm\epsffile{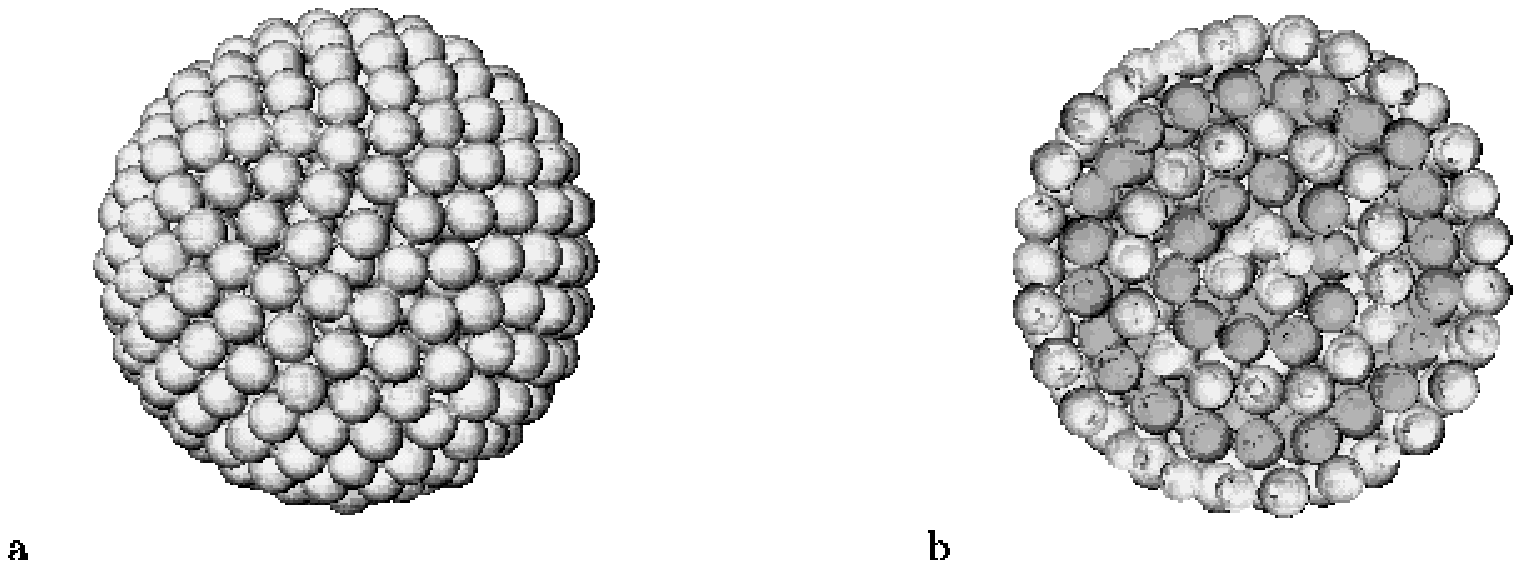}}
\vskip -0cm
\caption{A configuration with 500 points. On the left are all the points realized using 
the same method as in fig.~\ref{balls}. On the right is a slice through the centre with
 alternate shells having different colours.
\goodbreak {\em See 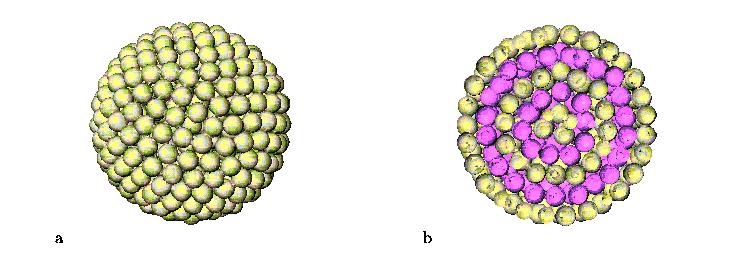 for a colour version of this figure.}}
\label{fig:500}
\end{center}
\end{figure}

\begin{figure} 
\begin{center}
\leavevmode
\epsfxsize=12cm\epsffile{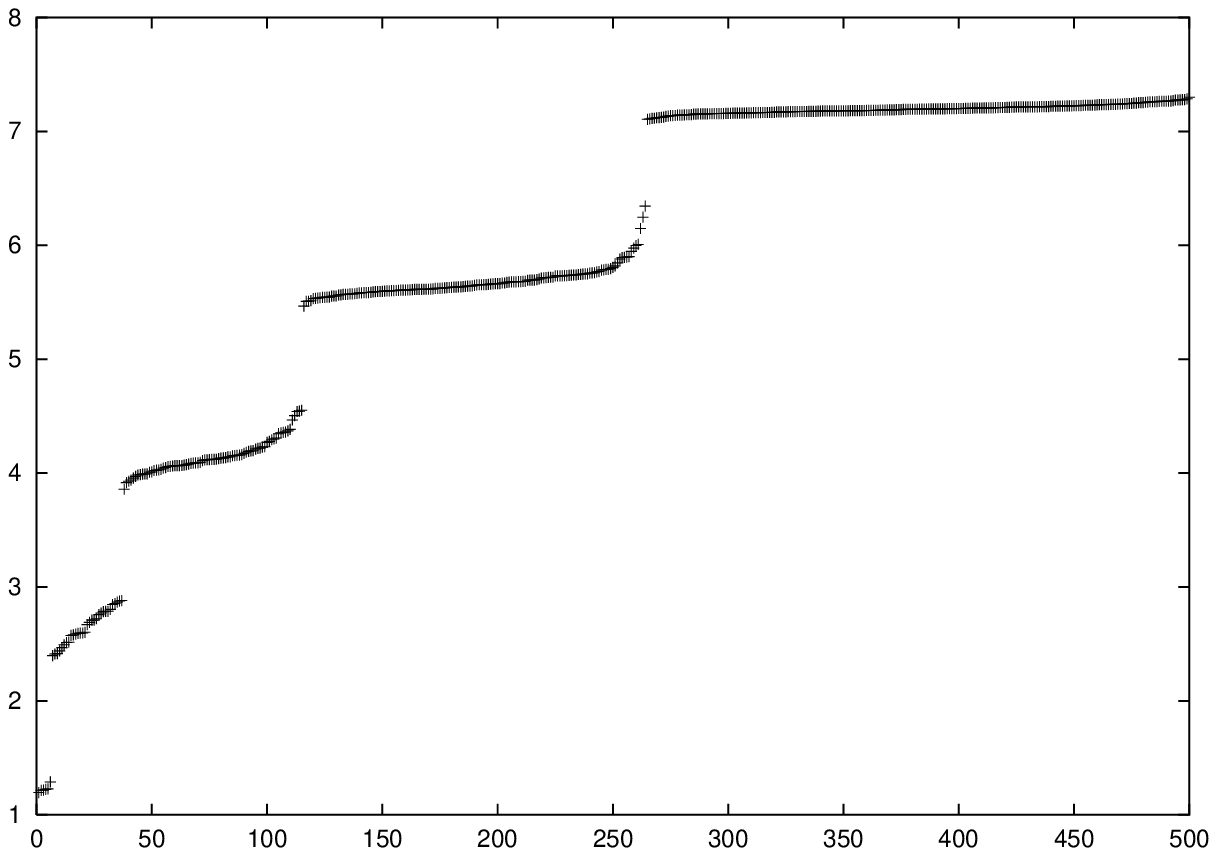}
\caption{The distance of each point from the origin for $N=500$}
\label{fig:500shells}
\end{center}
\end{figure}

To illustrate the discussion of the previous Section,
 here we will consider a specific example with $N=500$ points. 
The configuration may not necessarily be that of minimal energy, but 
\hbox{$V=27903.2=1.00018B_{500}$} where 
\hbox{$B_N=\textstyle{9\over 10}N(N^{2\over 3}-1)$}
denotes the lower bound given by (\ref{lbound}). 
The structure of the solution is illustrated in fig.~\ref{fig:500} and 
fig.~\ref{fig:500shells}. Fig.~\ref{fig:500} is created in a similar
 way to fig.~\ref{balls} by surrounding each point by a sphere of 
diameter $d=1.65$. On the right is a slice through the centre of the
 figure on the left with the spheres in different shells coloured differently.
 It is clear that there are 5 shells. This is further illustrated by
 fig.~\ref{fig:500shells} where the distance from the origin of each point is plotted. 
Notice that the shells are distinct in that there are obvious gaps between them.

\begin{figure} 
\begin{center}
\leavevmode
\epsfxsize=12cm\epsffile{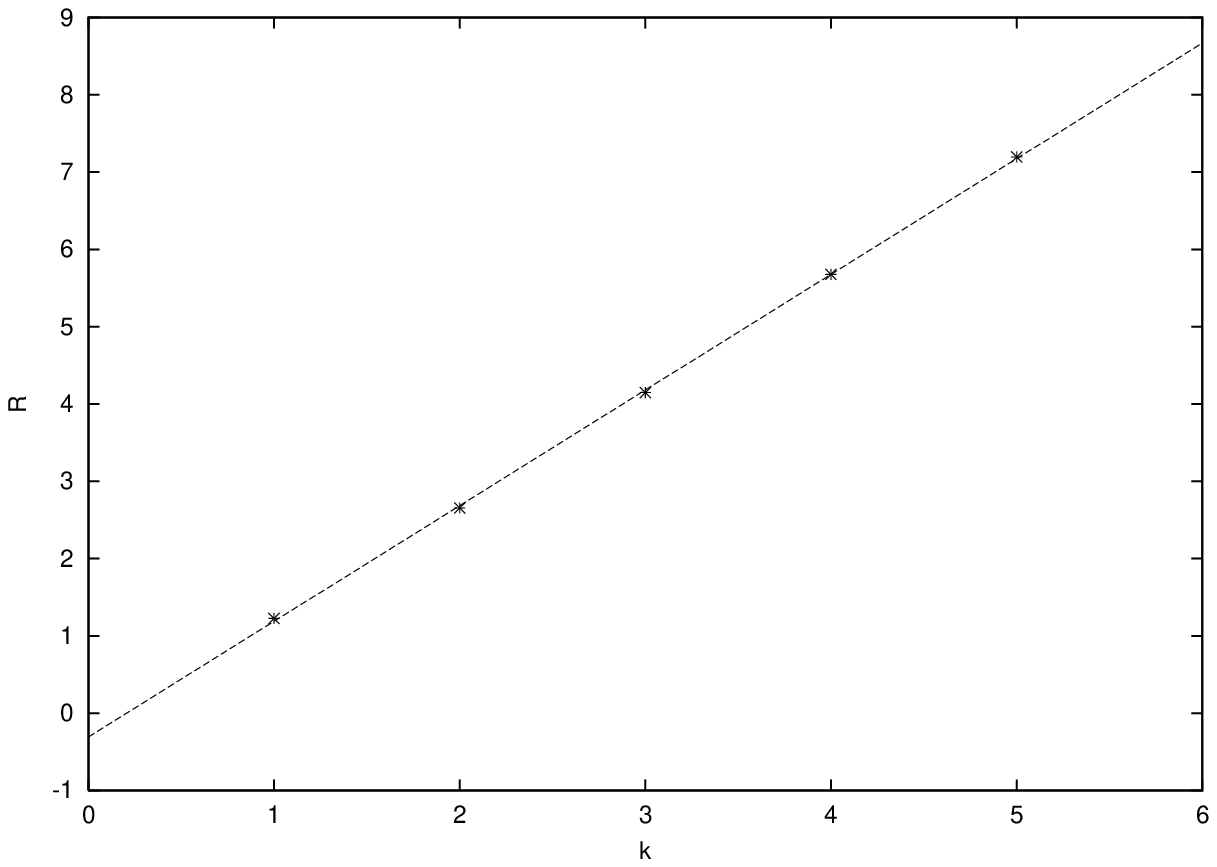}
\caption{The shell radii $R_k$ together with a linear fit, for an $N=500$
 configuration containing 5 shells.}
\label{fit}
\end{center}
\end{figure}

Table~\ref{example} lists the values of $\Delta N_k$, $R_k$ and $\sigma_k$
 for each of the shells. Included also is $\delta R_k$, the standard deviation 
of the radii of the particles in each shell, which shows that the inner shells
 are much less localized  than the outer crust. This can also be seen from
fig.~\ref{fig:500shells}.
Except for the innermost shell $\sigma\approx 4.5.$
The data values for $R_k$ are plotted in fig.~\ref{fit}
together with the linear fit $R_k=kR_{\rm c}-R_0$ 
using the values 
$R_{\rm c}= 1.5$ and $R_0=0.31.$
 Using these values, and the fact that there are
 5 shells, the final formula in (\ref{roughv}) gives
the estimate $V\approx 27315$, illustrating that this approximation
is accurate to within a few percent.

\subsection{Rough estimates} 

In this Section we describe how to obtain some rough estimates of
the inter-particle distance, the surface density, and the number of particles
in each shell.

A regular tetrahedron has a height $\sqrt{3}/2$
times the length of a side. Thus an estimate
for the inter-particle distance
is
\ben
d \approx { 2\over \sqrt{3} } (R_k-R_{k-1})
\approx {2R_{\rm c}}/{\sqrt{3}}
\label{crude1}.
\een
Using the earlier value of $R_{\rm c}=1.5$ yields
\ben
d \approx 1.73,
\een
which is not out of line with the absence of inter-particle forces
at very small and very large separations. 

To get a handle on the surface density we note that the closest
packing for circles on the plane (a problem originally tackled by
Kepler and by Harriot) is attained for hexagonal packing for which
the surface packing ratio is
\ben
\zeta = { \pi \over 2 \sqrt{3}} =0.9068996\dots
\een

A rough estimate for the number of spheres of diameter
$d$ that can be packed in a sphere
of radius $R_k$ is thus
\ben
\Delta N_k \approx {4 \pi R^2 _k \zeta \over \pi ({d \over 2})^2 }.
\label{crude2}
\een
Thus we get the estimate (better an upper bound)
\ben
4 \pi \sigma \approx { \Delta N_k \over R^2_k} \approx {16 \zeta \over d^2}.
\een
Substituting $d\approx 1.73$ yields
\ben
4 \pi \sigma_k \approx 4.85,
\een
which is certainly larger than 4.5 but not enormously so.

We can obtain a crude over-estimate for the number of particles in each shell
by replacing $R_k$ in (\ref{crude2}) by the approximation $R_k\approx kR_{\rm c}$,
where we have neglected the negative constant term in our earlier linear fit 
(the source of the over-estimation). Then using the final relation in (\ref{crude1})
we arrive at
\ben
\Delta N_k \approx 2 \pi \sqrt{3} k ^2\approx 10.88 k^2.
\een
Taking the integer part of this expression produces the values
$\Delta N_1=10, \Delta N_2=43, \Delta N_3=97, \Delta N_4=174, \Delta N_5=272$
which should be compared with those in Table~\ref{example}.
We see that these numbers are indeed over-estimates but give reasonable ball-park
values.

\section{Case III : large numbers of points}\news

\subsection{Statistical results: close packing and hard sphere model}
\begin{figure} 
\begin{center}
\leavevmode
\epsfxsize=12cm\epsffile{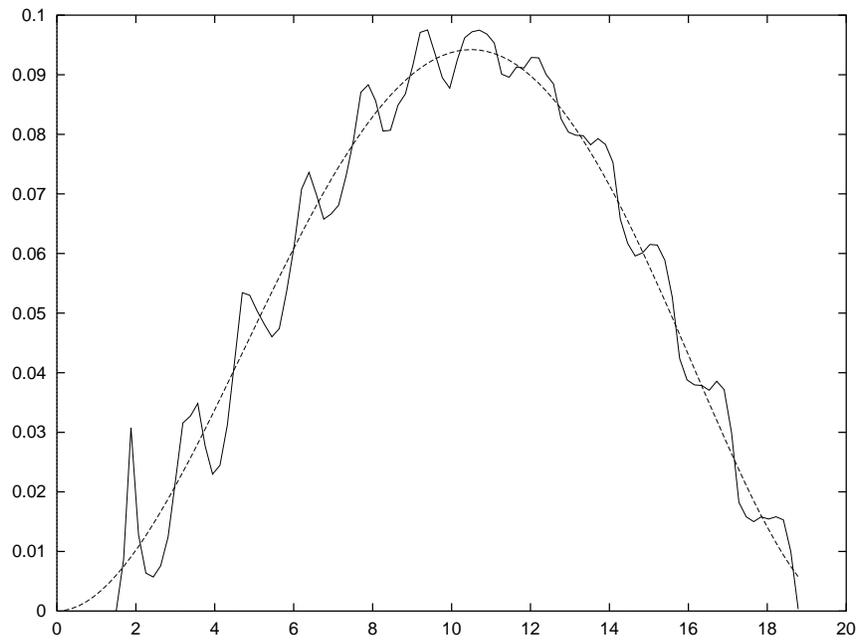}
\caption{The two-point separation distribution 
for 1000 points (solid line) and the Williamson distribution
(dashed line).}
\label{prob1000}
\end{center}
\end{figure}

The following statistical results are based on the lowest energy configuration
of 1000 points that we were able to compute. This configuration has
an energy $V=1.000103\,B_{1000},$ where 
$B_N={ 9 \over 10} N (N^{2 \over 3}-1)$ denotes the lower bound
given by (\ref{lbound}). Thus, although we are not able to claim that
this is the global minimum energy configuration, its energy is clearly very
close to that of the global minimum because of its small deviation from
the lower bound.
\begin{figure} 
\begin{center}
\leavevmode
\epsfxsize=12cm\epsffile{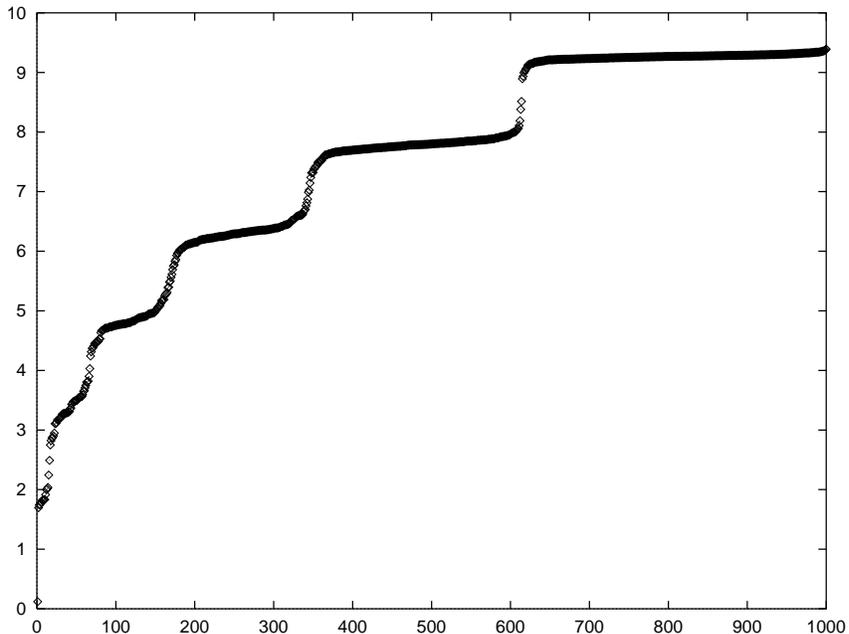}
\caption{The distance of each point from the origin for $N=1000.$}
\label{dist1000}
\end{center}
\end{figure} 
In fig.~\ref{dist1000} we plot the distance of each point from the origin for
$N=1000.$ This plot demonstrates that for 1000 points there is still
a shell-like structure, associated with the visible steps, but
the distinction between the shells is now quite blurred.
\begin{figure} 
\begin{center}
\leavevmode
\epsfxsize=12cm\epsffile{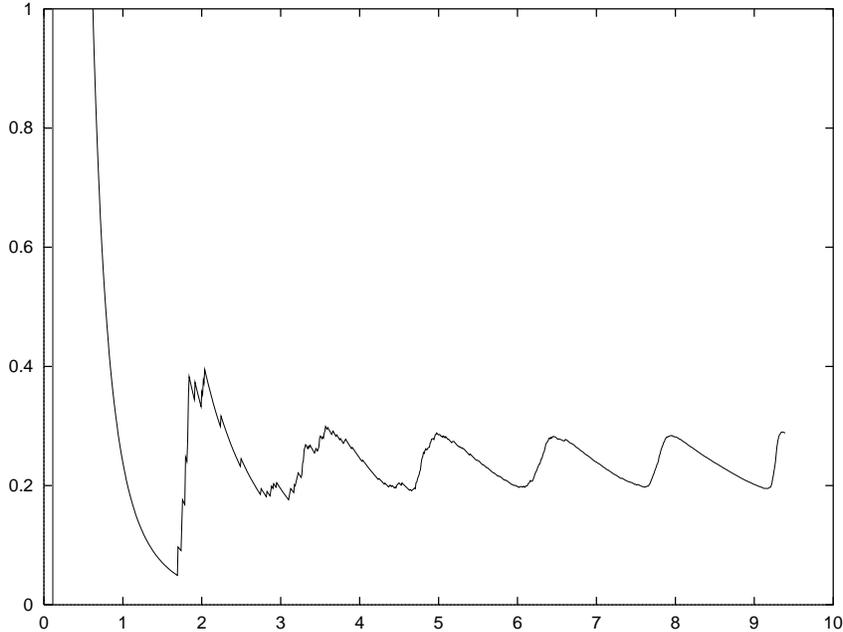}
\caption{The density as a function of radial distance for $N=1000.$}
\label{rho1000}
\end{center}
\end{figure} 
Fig.~\ref{rho1000} displays the density as a function of radial distance for
this configuration. A fairly constant amplitude oscillation around
the predicted  constant density ${3}/{4\pi}=0.2387..$
 suggests that the shells
are merging to form a uniformly distributed continuum. Further
evidence in support of this comes from computing the two-point
separation probability distribution and comparing with the
Williamson probability density (\ref{Willdist}). The results are
presented in fig.~\ref{prob1000}. The solid line is the numerically
computed separation distribution and the dashed line is the
Williamson probability density with $R=(N-1)^{1/3}$ and $N=1000.$
A convergence towards a uniform distribution is clearly suggested
by the data. Computing the average separation yields 
$\bar r=1.0251\,R$ which is again in good agreement with the
analytic result given by  (\ref{Willmean}).

\begin{figure}
\begin{center}
\leavevmode
\epsfxsize=12cm\epsffile{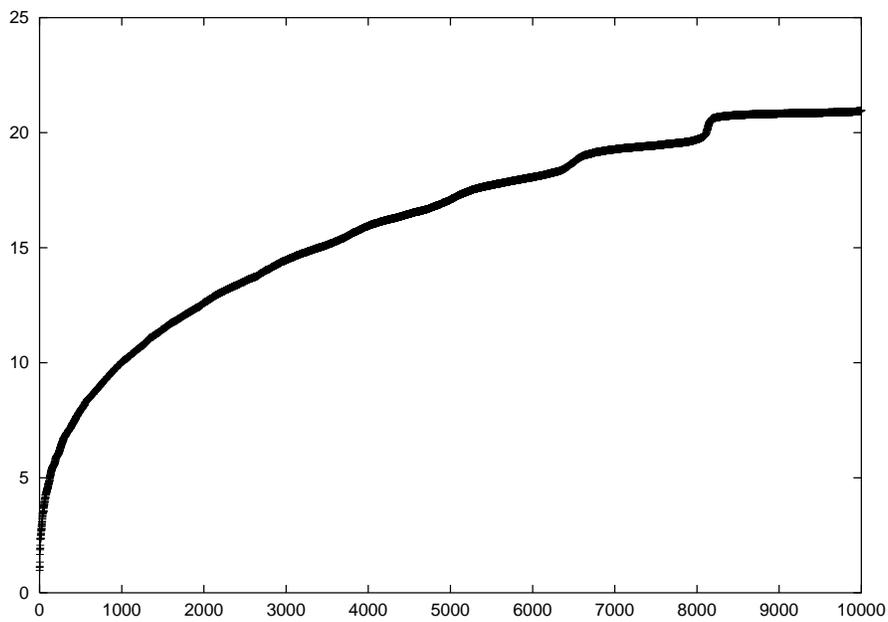}
\caption{The distance of each point from the origin for $N=10000.$}
\label{dist10000}
\end{center}
\end{figure}

\begin{figure}
\begin{center}
\leavevmode
\epsfxsize=12cm\epsffile{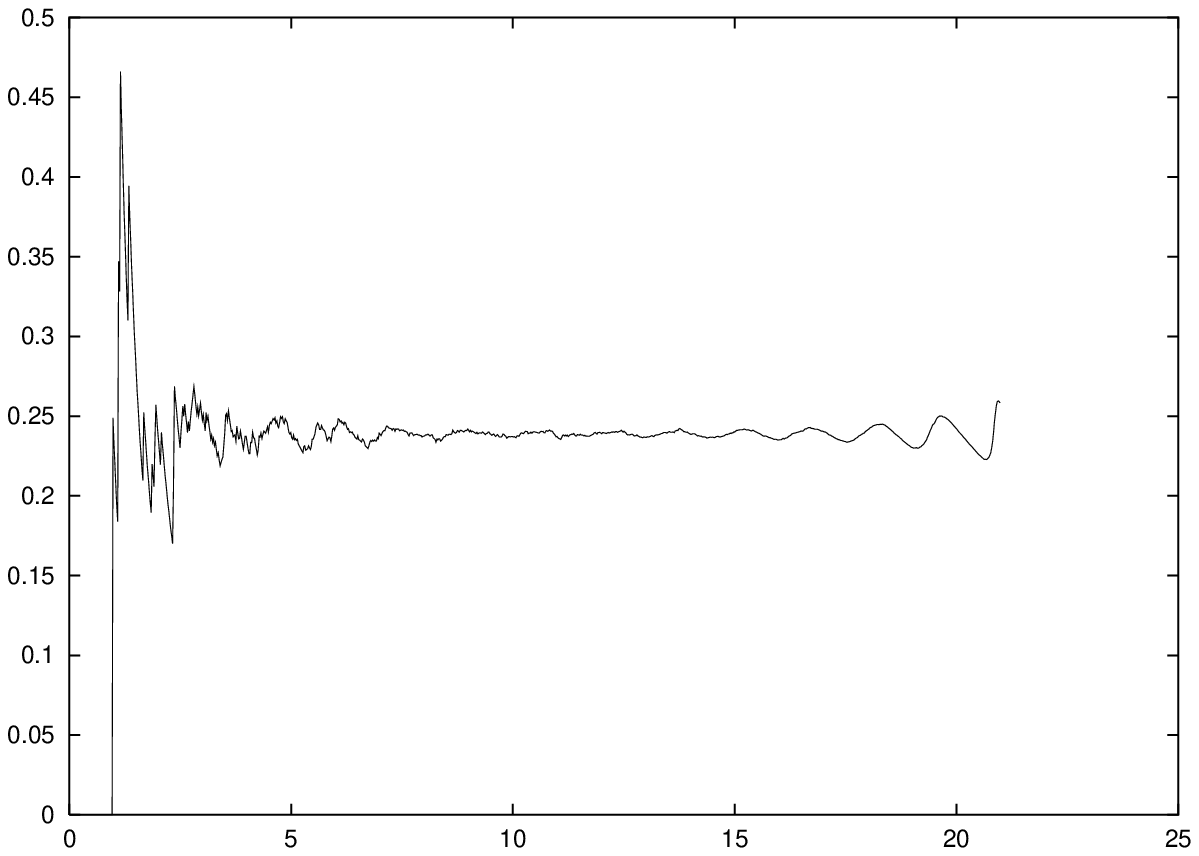}
\caption{The density as a function of radial distance for $N=10000.$}
\label{rho10000}
\end{center}
\end{figure}

\begin{figure}
\begin{center}
\leavevmode
\epsfxsize=12cm\epsffile{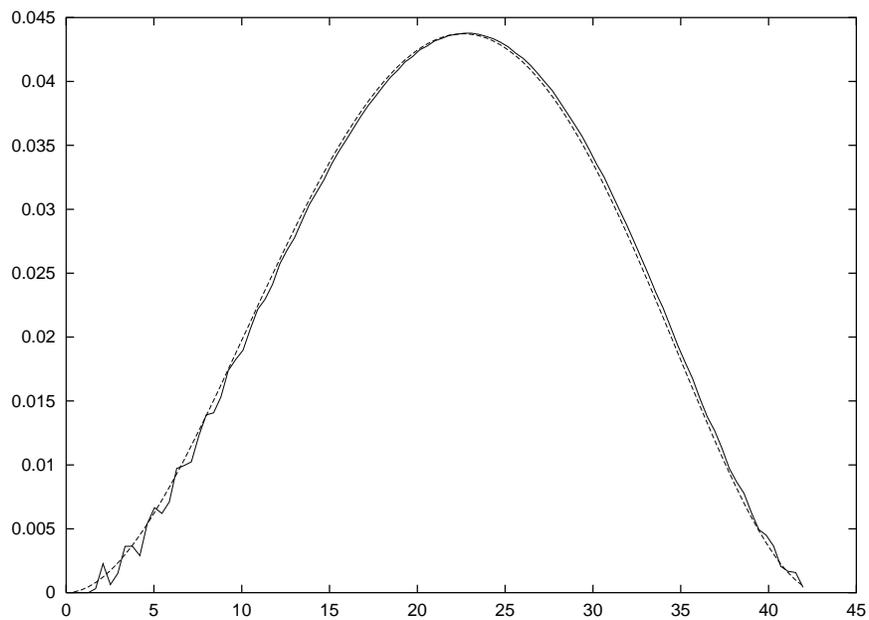}
\caption{The two-point separation distribution 
for 10000 points (solid line) and the Williamson distribution
(dashed line).}
\label{prob10000}
\end{center}
\end{figure}

To investigate the large $N$ limit further we compute the quantities
discussed above for $N=10000.$ The configuration we computed in this case
has energy $V=1.000022\,B_{10000},$ so again it is close to the
global minimum value.
In fig.~\ref{dist10000} we plot the distance from the origin of the
10000 points. In this case the individual shells have merged into
a continuum distribution, except for a crust layer near the edge
of the distribution where small steps can still be seen.
In fig.~\ref{rho10000} we display the density for this configuration, which
is now almost constant at the expected value ${3}/{4\pi}$ over a large
range. In fig.~\ref{prob10000} we compare the two-point separation
distribution (solid line) with the Williamson distribution (dashed line)
and find a remarkable agreement. The average separation computed from
our data is $\bar r=1.0277\,R,$ again a close fit to the analytic value.

\begin{figure}
\begin{center}
\leavevmode
\epsfxsize=12cm\epsffile{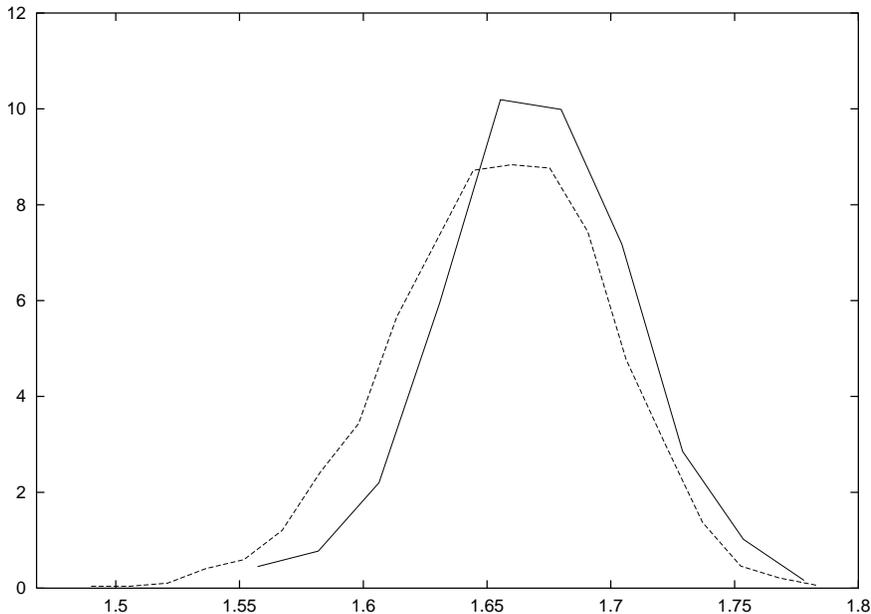}
\caption{The distribution of nearest neighbour separations for
$N=1000$ (solid line) and $N=10000$ (dashed line).}
\label{nn}
\end{center}
\end{figure}

In fig.~\ref{nn} we plot the distribution of nearest neighbour separations
for $N=1000$ (solid line) and $N=10000$ (dashed line). In both cases
all the nearest neighbour separations $r$ satisfy $1.48<r<1.80.$
The distributions are peaked around the value $r\approx 1.65$ which 
determined our earlier choice of the sphere packing diameter $d\approx 1.65.$

\begin{figure} 
\begin{center}
\leavevmode
\epsfxsize=8cm\epsffile{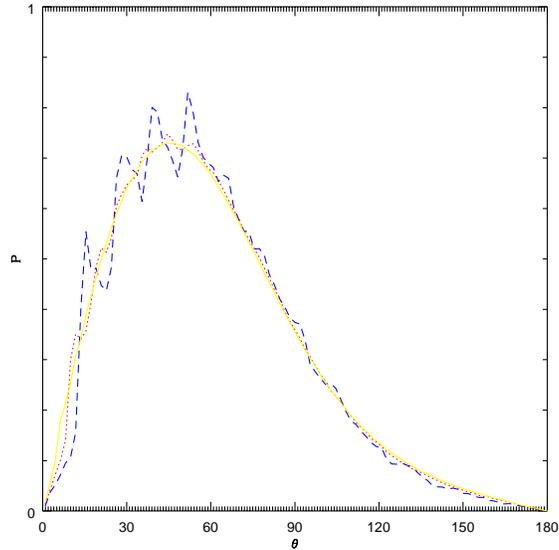}
\caption{The probability distribution of angles within triangles in configurations
 with $N=1000$ (solid line), $N=500$ (dotted line) and $N=75$ (dashed line).}
\label{tri1000}
\end{center}
\end{figure}

We have also computed the distribution of the angles within triangles formed from every
 triplet of points, analogous to a three-point function. This appears to be almost universal
 for all $N$ and is illustrated in fig.~\ref{tri1000} for $N=75$, $N=500$ and $N=1000$.
 The distributions for $N=500$ and $N=1000$ are almost identical, and only when $N=75$
 are there significant deviations from the universal distribution due to the effects of
 discreteness. We also computed the probability that the triangle was acute-angled which
 can be computed to be $33/70\approx 0.4714$ based on the Williamson distribution.
 For $N=75$ we computed this probability to be 0.4981 and it was 0.4685 and 0.4743 
for $N=500$ and $N=1000$
 respectively, all very close to the analytic value.

All the above 
 results are compatible
with a hard sphere model,
 similar to  Bernal's hard sphere model for liquids
in which one tries to pack a sphere of radius $R$ with $N$
impenetrable spheres of diameter $d.$

\begin{figure}
\leavevmode
\centerline{\epsfxsize=6cm\epsffile{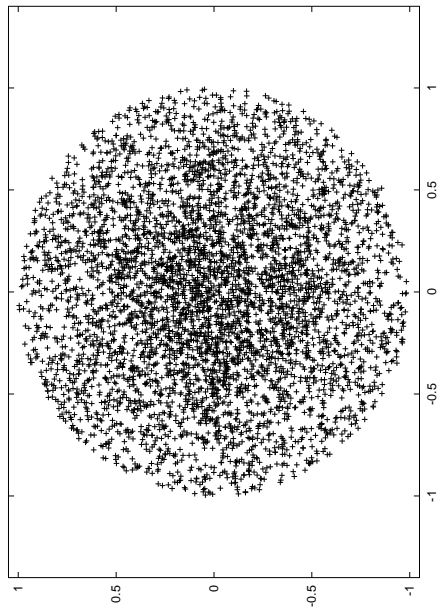}}
\caption{The distribution of nearest neighbour directions,
for $N=10000,$
as points on the unit sphere projected onto the unit disc.}
\label{direc}
\end{figure} 

\subsection{Crystallization and orientational order}\label{sec-cry}
Despite the wide-spread belief that in the limit of infinite numbers of
particles the minimum of the OCP model is given by a
Body Centred Cubic (BCC) 
 crystal, our preliminary results for up to 10,000
particles appear to show no sign of crystallization,
nor orientational order. 
To demonstrate this we compute, for each point, the direction of
its nearest neighbour. This gives a set of $N$ points on the unit sphere.
In fig.~\ref{direc} we display,
by stereographic projection onto the unit disc in the plane,
 the points obtained this way which lie
in the northern hemisphere (a similar plot is obtained for
the southern hemisphere).
After accounting for the slight distortion produced by the
stereographic projection we see that these points are essentially
distributed uniformly on the unit sphere. This indicates that there
is no orientational order or crystal structure. We have checked that these
results are not confused by any kind of a crust distribution by
confirming that a similar picture is obtained by computing only with a
central core of the configuration.

Of course one may always argue that
our numerical method has simply not found the global minimum energy
configuration, and hence we do not observe a crystal structure. It is
impossible to rule out this possibility, though there are a couple
of comments to be made which relate to this issue. The first is that
using the same numerical codes we have studied a two-dimensional version of
this problem, and found that a crystal structure does emerge and that
it is numerically easy to find and display. These results will be
presented elsewhere and tend to suggest that our codes should be
capable of finding a crystal structure if it is truly preferred.
The second point is that our numerical algorithms are based on
physical processes such as thermal fluctuations, so that even if we
have not found the global minima then these non-crystalline local
minima should still be of physical relevance.

\subsection{The packing fraction}

The packing fraction $\eta$ of $N$ spheres of radius $a$
confined to a volume $A$ is defined by
\ben
\eta= {4 \pi N a^3 \over 3A}.
\een
In our case
\ben
\eta = Nd^3/8 R^3.
\een
If we assume the earlier relation that $R=(N-1)^{ 1\over 3}$,
then for large $N$
\ben
\eta \approx \Big(\frac{d}{2}\Big)^3.
\een
Substituting $d=1.65$ gives $\eta=0.56.$

It is now known that hexagonal close packing (HCP) or face
 centred cubic close packing
(FCC)
have the densest possible value $\eta_{FCC}=0.74.$
Body centred cubic (BCC) has $\eta_{BCC}=0.68$, and simple cubic packing
(SCC) even smaller, $\eta_{SPC}=0.52$.

Numerical and experimental data used by
liquid theorists give mean values ${\bar \eta}_{RCP}=0.64$~\cite{ziman},
although the precise definition of random close packing (RCP) seems uncertain
(see \cite{TTD} for a recent discussion of this issue).
Nevertheless, the value of $\eta$ that we have obtained 
is in reasonable agreement with this one.

\section{Discussion and conclusions}\news
\label{sec-conc}

By use of numerical algorithms we have investigated in detail  central configurations
 where the interaction force is that of an inverse square  
law and the masses (charges)
of  all
 the particles are equal. We find that for low values of $N$ the configurations are
 generally convex deltrahedra which gives way to a multi-shell structure for $N>12$.
 As $N$
 increases the number of shells increases and eventually the configuration tends
 towards having a constant density. The two-point probability distribution and also the
 probability of acute angle triangles agree to a high degree with those of a uniform
 distribution. The distribution of nearest neighbours is sharply peaked suggesting that
 each particle can be approximated by a sphere of diameter $d\approx 1.65$ and we have
 found, at this stage, no evidence for long-range orientational order in contrast to the
 situation in 2-dimensions, which we shall present elsewhere. 
It still remains an open question as to whether crystallization occurs, and
the possibility remains that for large values of $N$  either we may not have 
 found a minimum sufficiently close to the global one, or 
that we have not probed sufficiently large values of $N$. 
These aspects are currently under further investigation.

\begin{figure}
\begin{center}
\leavevmode
\centerline{\epsfxsize=10cm\epsffile{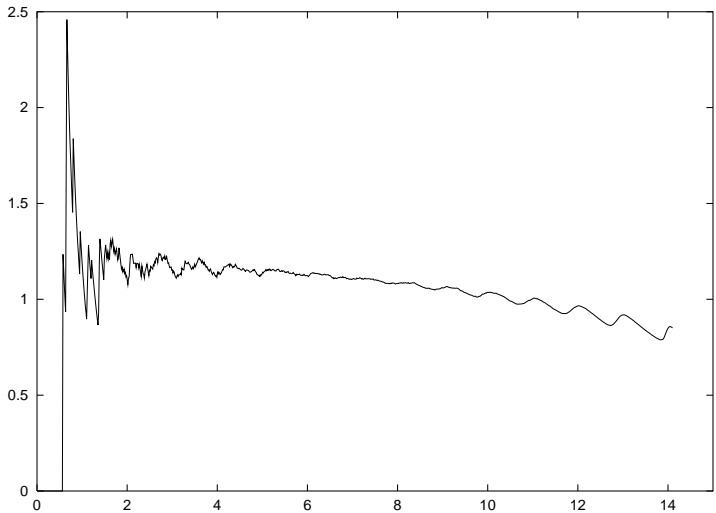}}
\caption{The density as a function of radial distance 
in a model where the interaction is generated by an
 inverse cube force with $N=10000$. Notice that the density is not
 constant on the outer extremities.}
\label{dendif}
\end{center}
\end{figure} 

The specific types of central configurations that we have computed are
examples for just one of a large set of models. As we have explained, the interaction
potential we have studied has a number of special properties, and we should 
note that different force laws will lead
 to very different results. To illustrate this we have included fig.~\ref{dendif} which
 shows the density distribution as a function of radial distance for particles with
 $N=10000$ when the interaction force is
 an inverse cube law. Clearly in this case there is a decreasing trend in the density
 with increasing radius, rather than the approach to uniform density that we have
encountered so far in this paper. If the power in the interaction force is
further increased then this downward trend becomes even more apparent.

\begin{figure}
\begin{center}
\leavevmode
\ \vskip -0cm
\centerline{\epsfxsize=12cm\epsffile{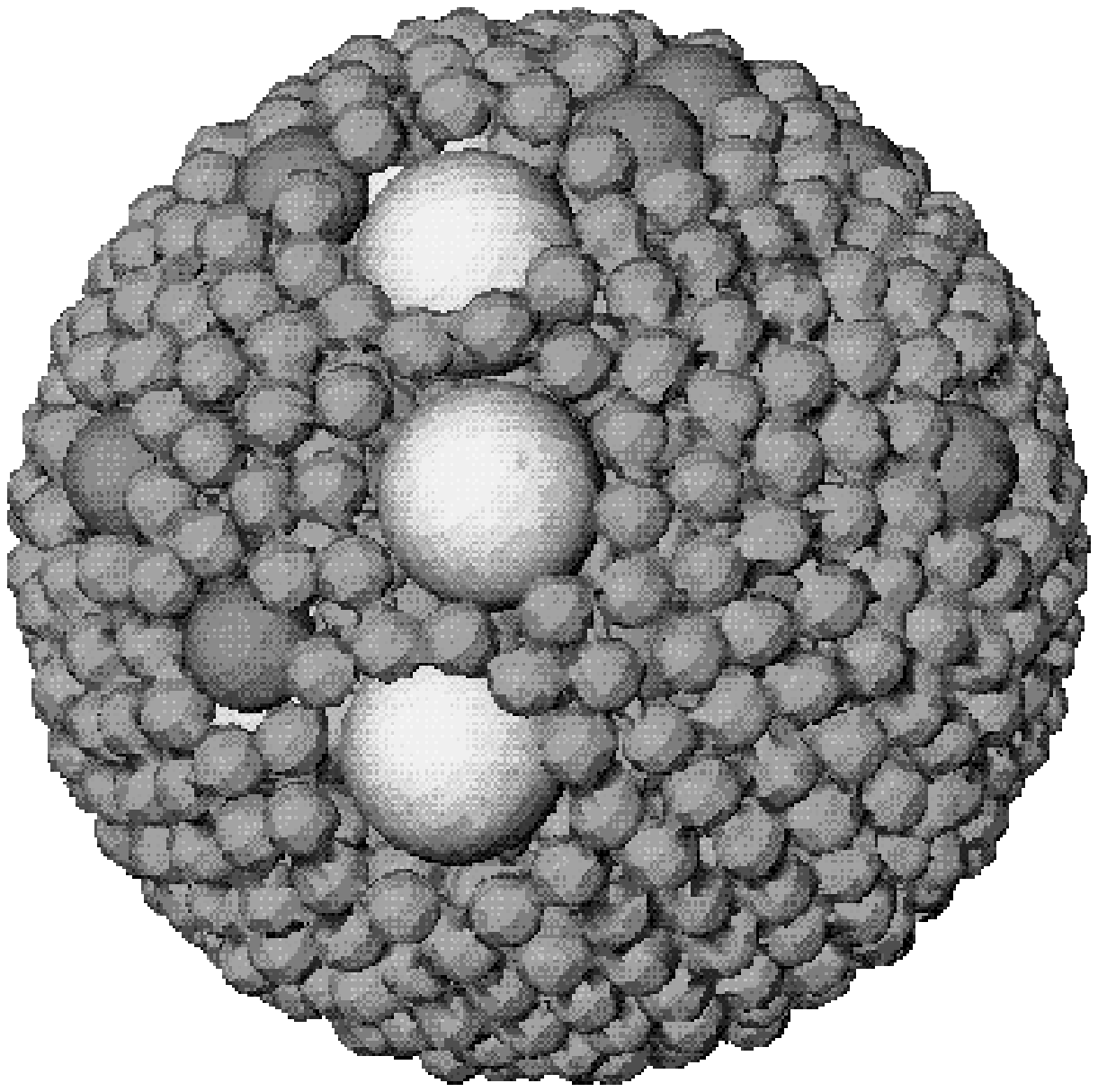}}
\caption{The distribution of 124 points, 100 of which have $m=1$, 20 have $m=5$ and 4
 have $m=25$. Each point with $m=1$ is represented by a sphere of diameter $d=1.65$
 and the others by spheres with a diameter related to their mass by $d\propto m^{1/3}$. 
\goodbreak {\em See 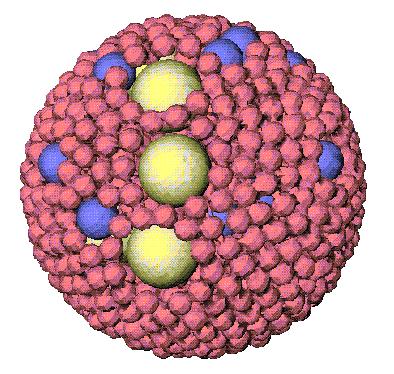 for a colour version of this figure.}
}
\label{diffmass}
\end{center}
\end{figure} 

Another interesting possibility is to consider situations in which
the particles have different masses.
 Using the intuition that each of the particles can be represented by a sphere, 
our earlier analysis suggests that the diameter of this sphere should be
 taken to be proportional to
$m^{1/3}$ and indeed we find this to be the case.
 This is illustrated in fig.~\ref{diffmass}, where the spheres can be seen to fit
snugly together using the above prescription of taking the volume of the sphere
proportional to the mass of the particle.

Clearly the current work is only the tip of the iceberg in terms of the full generality
 of the concepts involved in central configurations, but we believe it represents a good
 starting point for further work. Investigations  into different power laws for the
 interactions,  different mass distributions and the all important question of 
whether crystallization occurs in these types of models are all underway.\\

\section*{Acknowledgements}
 We acknowledge advanced fellowships from PPARC (RAB) and EPSRC (PMS).
We thank Mike Moore for useful discussions.\\

\noindent{\em Note Added}

After the submission of this paper there appeared \cite{Totsuji}
who treat larger values of $N$ than we do here. They find numerical
evidence for a transition at $N=N_c$ to a BCC structure, where $1.1 \times
10 ^4 < N_c < 1.5 \times  10 ^4$. From this paper we became aware of
earlier relevant papers including \cite{Rafac, Tsuruta} whose results for
small values of $N$ have considerable  overlap with our own.
Where comparisons are possible, we find good agreement both qualitatively
and quantitatively. Another relevant reference is \cite{CedoLlibre} who
list all symmetric (but not necessarily stable) solutions.
Again, where comparisons are possible, their and our results agree.\\

\end{document}